\documentclass[conference]{IEEEtran}
\IEEEoverridecommandlockouts
\usepackage{cite}
\usepackage[ruled,linesnumbered]{algorithm2e}
\usepackage{amsmath,amssymb,amsfonts}
\usepackage{algorithmic}
\usepackage{graphicx}
\usepackage{textcomp}
\usepackage{graphicx}
\usepackage{subcaption}  
\usepackage{xcolor}
\usepackage{float}
\usepackage[hang,flushmargin]{footmisc}
\newtheorem{proposition}{\underline{Proposition}}

\newtheorem{lemma}{\underline{Lemma}}
\def\BibTeX{{\rm B\kern-.05em{\sc i\kern-.025em b}\kern-.08em
		T\kern-.1667em\lower.7ex\hbox{E}\kern-.125emX}}
\newcommand{\mv}[1]{\mbox{\boldmath{$ #1 $}}}
\usepackage[T1]{fontenc}
\usepackage{times}
\usepackage{times}
\usepackage[T1]{fontenc}

\begin{document}
	\title{Base Station Placement Optimization for Networked Sensing Exploiting Target Location Distribution\\
	}
	\author{\IEEEauthorblockN{Kaiyue Hou and Shuowen Zhang}
		\IEEEauthorblockA{{Department of Electrical and Electronic Engineering, The Hong Kong Polytechnic University} \\
			E-mails: kaiyue.hou@connect.polyu.hk, shuowen.zhang@polyu.edu.hk}
	}
	
	\maketitle
	
	\begin{abstract}
	This paper studies a networked sensing system with multiple base stations (BSs), which collaboratively sense the \emph{unknown} and \emph{random} three-dimensional (3D) location of a target based on the target-reflected echo signals received at the BSs. Considering a practical scenario where the target location distribution is known \emph{a priori} for exploitation, we aim to design the placement of the multiple BSs to optimize the networked sensing performance. Firstly, we characterize the \emph{posterior Cram\'er-Rao bound (PCRB)} of the mean-squared error (MSE) in sensing the target's 3D location. Despite its complex form under networked sensing, we derive its \emph{closed-form} expression in terms of the BS locations. Next, we formulate the BS placement optimization problem to minimize the sensing PCRB, which is non-convex and difficult to solve. By leveraging a series of equivalent transformations and the iterative inner approximation method, we devise an algorithm with polynomial-time complexity which is guaranteed to converge to a solution satisfying the \emph{Karush-Kuhn Tucker (KKT) conditions} of the problem. Numerical results show that the proposed placement design significantly outperforms various benchmark designs.
	\end{abstract}
	\vspace{-2mm}
	\section{Introduction}
	\vspace{-2mm}
	The sixth-generation (6G) cellular networks will support sensing as a critical new function \cite{liu2022integrated}. To this end, various classic communication devices can serve as sensing anchors, including the cellular base stations (BSs), access points (APs), intelligent reflecting surfaces, or even user equipment \cite{CM}. Moreover, multiple anchors can collaborate to perform \emph{networked sensing} \cite{shi2022device}, which unlocks significant performance gains over traditional sensing systems such as radar systems.
	
	The performance of sensing is critically dependent on the placement (i.e., locations) of the sensing anchors. For instance, for range-based localization via estimating the time-of-arrivals (ToAs) of the target-reflected paths, shortening the distance between anchors and the target will enhance the path power, thereby increasing the ToA estimation accuracy. Assuming that the target's location is deterministic and known, \cite{godrich2010target} studied the anchor placement optimization to minimize the Cram\'er-Rao bound (CRB) of the sensing mean-squared error (MSE);  \cite{nguyen2016optimal, fatima2024optimal} pursued the optimal system layouts for ToA-based localization; \cite{aubry2023robust} further proposed robust placement designs by optimizing the average and worst-case CRB.
	
	This paper aims to study a novel and practical scenario where the target's location to be sensed is \emph{unknown} and \emph{random}, while its \emph{distribution} is heterogeneous and known for exploitation based on historic data or target appearance pattern \cite{xu2023mimo,xu2024mimo}. In this case, \emph{posterior Cram\'er-Rao bound (PCRB)} \cite{bibTrees} is a suitable performance metric, which is a lower bound of the MSE with any unbiased estimator exploiting prior distribution information, and is only dependent on the distribution instead of the actual value of the target location. Although prior studies have demonstrated the effectiveness of PCRB in guiding beamforming design \cite{xu2023mimo,xu2024mimo,Hou_JSAC}, how to design the anchor placement based on PCRB is still an unaddressed and challenging problem, as the PCRB is a more complex function with respect to the anchor locations than beamforming, especially in networked sensing systems.
	
	\begin{figure}
		\centering
		\vspace{-1mm}
		\includegraphics[width=0.83\linewidth]{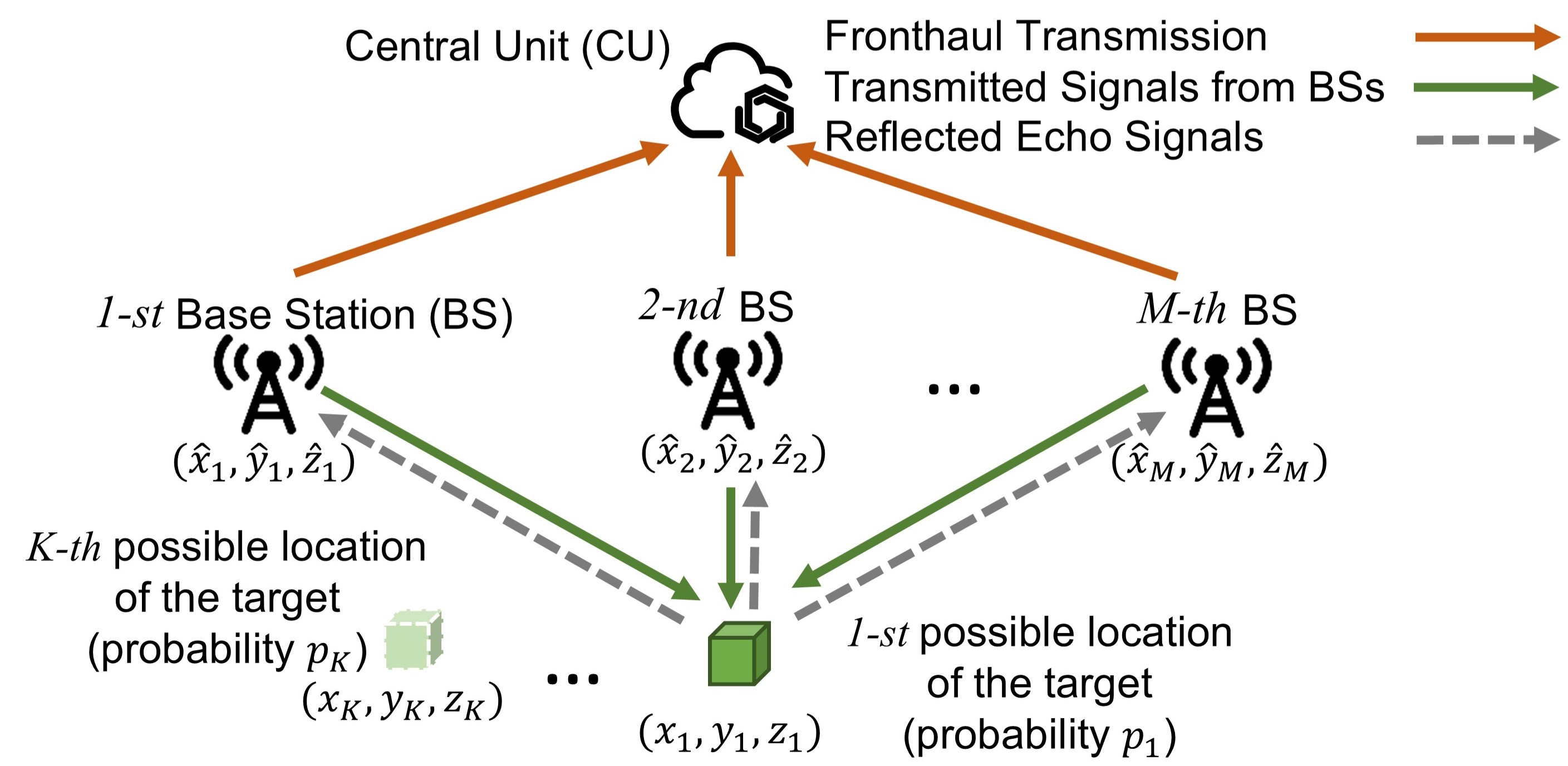}
		\setlength{\abovecaptionskip}{-0.1cm}\vspace{2mm}
		\caption{Illustration of a multi-BS networked sensing system.}\vspace{-2mm}
		\label{Fig1_system}
	\end{figure}

	In this paper, we aim to fill the above research gap by focusing on a multi-BS networked sensing system under orthogonal frequency division multiplexing (OFDM). The BSs are connected to a central unit (CU) and collaboratively sense the \emph{unknown} and \emph{random} three-dimensional (3D) location of a target based on its reflected echo signals, as illustrated in Fig. \ref{Fig1_system}. Based on the known prior distribution information of the target's location, we first characterize the PCRB of the sensing MSE in \emph{closed form} with respect to the locations of the multiple BSs, which are then jointly optimized to minimize the PCRB. Despite the non-convexity of this challenging problem, we devise an \emph{iterative inner approximation} based algorithm that is guaranteed to converge to a \emph{Karush-Kuhn Tucker (KKT) point} within polynomial time. It is shown via numerical results that our proposed placement design achieves superior sensing performance over various benchmarks.
	\vspace{-2mm}
	\section{System Model}
	\vspace{-1mm}
	We consider an OFDM-based multi-BS networked sensing system to estimate the \emph{unknown} and \emph{random} 3D location of a point target denoted by $\mv{u}=[x,y,z]^T$ in meters (m). The distribution of the target's location is assumed to be known \emph{a priori} based on historic data or target appearance pattern \cite{xu2023mimo}, \cite{xu2024mimo}. As illustrated in Fig. \ref{Fig1_system}, the target has $K>1$ possible locations denoted by set $\mathcal{K}=\{1,\dots,K\}$. Each possible location is given by $[x_k,y_k,z_k]^T$ with a probability mass of $p_k\!\in \!(0,1)$, where $\sum_{k=1}^K\!p_k\!=\!1$. The probability mass function (PMF) of the target's 3D location is thus expressed as 
	\begin{align}\label{P_target_BS}
		\hspace{-2mm}{p}_{{X,Y,Z}}(x,y,z)\!\! =\!\! \left\{\begin{matrix}
			p_k, &\!\!\! {\rm{if}}\ \!x\!=\!x_k,\! \ \!y\!=\!y_k,\! \ \!z\!=\!z_k, \ \!k\!\in\!\mathcal{K},\! \\
			0\ , &\hspace{-2mm} {\rm{otherwise}}.\qquad\qquad\qquad\qquad \\
		\end{matrix}\right.
	\end{align}

	The system consists of $M\geq 4$ BSs denoted by set $\mathcal{M}=\{1,\dots,M\}$, which are connected to a CU via fronthaul links. Denote $N>1$ as the total number of OFDM sub-carriers, and $\mathcal{N}_m\subset\{1,\dots,N\}$ as the set of sub-carriers allocated to each $m$-th BS, with $\mathcal{N}_m\cap\mathcal{N}_n=\emptyset$, $\forall m\neq n$, i.e., the BSs are allocated with non-overlapping sub-carriers. Each BS utilizes one dedicated transmit antenna to send downlink wireless signals, and one dedicated receive antenna to receive the echo signals reflected by the target over its assigned sub-carriers. The BSs then send the received signals to the CU via fronthaul links for networked sensing. We aim to design the placement of the $M$ BSs to optimize the networked sensing performance. Each $m$-th BS is assumed to have a fixed height of $\hat{z}_m$ (m) due to site requirements, and an adjustable horizontal location $[\hat{x}_m,\hat{y}_m]^T$ (m) which will be optimized.
	
	Denote $\Delta_f$ in Hz as the OFDM sub-carrier spacing. The overall system bandwidth is thus given by $B=N\Delta_f$ Hz. Let $\mv{s}_m\in \mathbb{C}^{|\mathcal{N}_m|\times 1}$ denote the frequency-domain OFDM symbols sent from BS $m$, where each entry represents a pilot/data symbol with unit average power. The distance between the target and each $m$-th BS is given by
	\begin{equation}\label{r_m}
		r_{m}= \sqrt{(\hat{x}_m-x)^2+(\hat{y}_m-y)^2+(\hat{z}_m-z)^2}.
	\end{equation}
	To draw fundamental insights into the BS placement optimization, we focus on the scenario where the target has line-of-sight (LoS) channels with the BSs. The propagation delay of the target-reflected round-trip signal path associated with BS $m$ is thus given by $\tau_{m} = \frac{2r_{m}}{c}$, where $c$ denotes the speed of light. Let $\alpha_m=\alpha_m^{\mathrm{R}}+j\alpha_m^{\mathrm{I}}\in \mathbb{C}$ denote the radar cross-section (RCS) coefficient of the target in its reflected path associated with the $m$-th BS, where $\alpha_m$'s are generally independent of each other, and can be assumed to follow the circularly symmetric complex Gaussian (CSCG) distribution with zero mean and variance $\sigma_\alpha^2$ \cite{skolnik1980introduction}, i.e., $\alpha_m\sim \mathcal{CN}(0,\sigma_\alpha^2)$. Let $\hat{p}_m={\frac{P}{|\mathcal{N}_m|}}$ denote the transmit power at each sub-carrier of BS $m$ with $P$ denoting the transmit power budget at each BS and equal power allocation among all sub-carriers assigned to the same BS, $\beta_0$ denote the reference channel power at distance $1$ m, $\beta_{m}\triangleq \frac{\beta_0}{r_m^2}\alpha_{m}\in\mathbb{C}$ denote the overall reflection coefficient of the BS $m$-target-BS $m$ reflected channel, and ${\mv{z}}_{m}\sim\mathcal{CN}({\mv{0}}, \sigma^2_z{\mv{I}}_{|\mathcal{N}_m|})$ denote the CSCG noise vector at the $m$-th BS receiver over the $|\mathcal{N}_m|$ sub-carriers, with $\sigma_z^2$ denoting the average noise power at each sub-carrier. The baseband equivalent  received signal at each $m$-th BS is then given by
	\begin{align}
		\mv{y}_m 
		= \sqrt{\hat{p}_m}\beta_m{\rm{diag}}(\mv{s}_{m})\mv{g}_m+\mv{z}_m, \  \forall m\in\mathcal{M},
	\end{align}
	where $\mv{g}_m=[e^{-j\frac{2\pi 2r_m}{\frac{c}{\mathcal{N}_m(1)\Delta_f}}},\dots, e^{-j\frac{2\pi 2r_m}{\frac{c}{\mathcal{N}_m(|\mathcal{N}_m|)\Delta_f}}}]^T=[e^{-j2\pi\mathcal{N}_m(1) \Delta_f \tau_{m}},\dots, e^{-j2\pi\mathcal{N}_m(|\mathcal{N}_m|) \Delta_f \tau_{m}}]^T\in\mathbb{C}^{|\mathcal{N}_m|\times 1}$, with $\mathcal{N}_m(n)$ denoting the $n$-th smallest element in the set $\mathcal{N}_{m}$. Each BS $m$ sends its received signal $\mv{y}_m$ to the CU.\footnote{To draw fundamental insights, we focus on an ideal scenario with infinite fronthaul capacity, while our framework is also applicable to the case with limited fronthaul capacity by performing sensing based on quantized $\mv{y}_m$'s.} The collection of received signals at the CU is thus given by
	\begin{align}
		{{\bar{\mv{y}} = \left[\mv{y}_1^H,\dots, \mv{y}_M^H\right]^H}}.
	\end{align} 
	Note that $\bar{\mv{y}}$ is a function of the target's 3D location $\mv{u}$, the RCS coefficients associated with the $M$ BSs denoted by  $\mv{\alpha} = [\alpha_{1},\dots, \alpha_{M}]^T\in \mathbb{C}^{M\times 1}$, and the locations of the $M$ BSs.
	
	The objective of networked sensing is to estimate $\mv{u}$ based on the \emph{observations} (measurements) in $\bar{\mv{y}}$ at the CU as well as the \emph{prior distribution information} in ${p}_{{X,Y,Z}}(x,y,z)$. It is worth noting that there are various practical methods to achieve this goal. For example, by estimating the propagation delays, $\tau_m$'s, of the target-reflected paths via OFDM channel estimation techniques, the range of the target with respect to each BS, $r_m$'s, can be obtained, which enables the CU to localize the target in the 3D space as the intersection of at least four spheres \cite{shi2022device}. Moreover, the prior distribution information can further enhance the estimation performance. 
	
	Since the exact MSE with each specific method is difficult to be expressed analytically, we adopt the \emph{PCRB} to characterize a global lower bound of the MSE with any unbiased estimator exploiting prior information. In the next section, we derive the PCRB as an explicit function of the BSs' locations, based on which the BSs' locations will be further optimized.
	
	\section{Networked Sensing Performance Characterization via PCRB}
	Note that besides the target's 3D location $\mv{u}$, the RCS coefficients in $\mv{\alpha}$ are also unknown parameters in $\bar{\mv{y}}$, thus need to be jointly estimated with $\mv{u}$. Let $\mv{\omega} = [\mv{u}^T,{\mv{\alpha}}^{{\rm{R}}^T}, {\mv{\alpha}}^{{\rm{I}}^T}]^T\in\mathbb{R}^{(3+2M)\times 1}$ denote the collection of unknown parameters, with $\mv{\alpha}^{\rm{R}} = \mathfrak{Re}\{\mv{\alpha}\}$ and $\mv{\alpha}^{\rm{I}} = \mathfrak{Im}\{\mv{\alpha}\}$. Since classic PCRB analysis is generally suitable for parameters with continuous and differentiable probability density functions (PDFs), we approximate the discrete PMF in (\ref{P_target_BS}) with a continuous and differentiable \emph{Gaussian mixture PDF} given by: 
	\begin{align}
		\hspace{-2.5mm}\!\bar{p}_{X,Y,Z}(x,y,z)\!=&\!\!\sum_{k=1}^{K}\!\frac{p_k}{(2\pi\sigma^2)^{\frac{3}{2}}}e^{-\frac{(x-x_k)^2+(y-y_k)^2+(z-z_k)^2}{2\sigma^2}}.\!\label{pxy}
	\end{align}
	Specifically, $\bar{p}_{X,Y,Z}(x,y,z)$ is the weighted summation of $K$ 3D Gaussian PDFs, each with mean $(x_k, y_k,z_k)$ and a small variance $\sigma^2$; moreover, $X$, $Y$, and $Z$ in $\bar{p}_{X,Y,Z}(x,y,z)$ are independent. As $\sigma^2$ approaches zero, the Gaussian mixture PDF asymptotically approaches the discrete PMF.\footnote{The efficacy of this approximation has been shown in \cite{Hou_JSAC}.} By noting that all parameters in $\mv{\omega}$ are independent and  $\alpha_m\sim \mathcal{CN}(0,\sigma_\alpha^2),\ \forall m\in \mathcal{M}$, the joint PDF of the parameters in $\mv{\omega}$ can be expressed as 
	$p_{\mv{w}}(\mv{\omega}) =\bar{p}_{X,Y,Z}(x,y,z)\prod_{m=1}^M \frac{1}{{\pi}\sigma^2_{\alpha}}e^{-\frac{1}{\sigma_{\alpha}^2}({\alpha}_m^{\mathrm{R}^2}+{\alpha}_m^{\mathrm{I}^2})}$.
	
	The posterior Fisher information matrix (PFIM) for the estimation of $\mv{\omega}$ consists of two parts as follows: 
	\begin{equation}
		\mv{F} = \mv{F}_{\rm{O}}+\mv{F}_{\rm{P}}.
	\end{equation}
	\vspace{-2mm}
	\subsection{Derivation of $\mv{F}_{\mathrm{O}}$}
	\vspace{-4mm}
	Firstly, $\mv{F}_{\rm{O}}=\mathbb{E}_{\bar{\mv{y}}, \mv{\omega}}\bigg[\frac{\partial{\rm{ln}}(f(\bar{\mv{y}}|\mv{\omega}))}{\partial \mv{\omega}}\bigg(\frac{\partial{\rm{ln}}(f(\bar{\mv{y}}|\mv{\omega}))}{\partial \mv{\omega}}\bigg)^{H}\bigg]\in\mathbb{R}^{(3+2M)\times (3+2M)}$ represents the PFIM extracted from observation, where $f(\bar{\mv{y}}|\mv{\omega})$ represents the conditional PDF of $\bar{\mv{y}}$ given $\mv{\omega}$. Note that a key difficulty in deriving $\mv{F}_{\mathrm{O}}$ lies in the complex relationship between $\mv{u}$ and $\bar{\mv{y}}$ through the $M$ BSs. To cross this hurdle, we introduce an auxiliary vector $\mv{\phi}=[\mv{r}^T,\mv{\alpha}^{{\rm{R}}^T},\mv{\alpha}^{{\rm{I}}^T}]^T\in\mathbb{R}^{3M\times1}$, where $\mv{r} = [r_{1}, \dots, r_{M}]^T$ contains the collection of  BS-target distances for all $M$ BSs defined in (\ref{r_m}). Notice that $\mv{r}$ and $\mv{u}$ (and consequently $\mv{\phi}$ and $\mv{w}$) have a unique one-to-one mapping for any feasible $\mv{r}$. Then, $\mv{F}_{\mathrm{O}}$ is expressed via the chain rule as
	\begin{align}
		&\mv{F}_{\rm{O}}\nonumber\\
		=&\mathbb{E}_{\bar{\mv{y}}, \mv{\omega}}\bigg[\bigg(\frac{\partial {\mv{\phi}}}{\partial \mv{\omega}}\bigg)^{H}\frac{\partial{\rm{ln}}( f(\bar{\mv{y}}|\mv{\phi}) )}{\partial \mv{\phi}}\bigg( \bigg(\frac{\partial {\mv{\phi}}}{\partial \mv{\omega}}\bigg)^{H}\frac{\partial{\rm{ln}}( f\left(\bar{\mv{y}}|\mv{\phi}\right))}{\partial \mv{\phi}}\bigg)^{H}\bigg]
		\nonumber\\
		%	=&\mathbb{E}_{\mv{u}}\bigg[\bigg(\frac{\partial {\mv{\phi}}}{\partial \mv{\omega}}\bigg)^{H}\mathbb{E}_{\bar{\mv{y}},\mv{\alpha}}\bigg[\frac{\partial{\rm{ln}}(f(\bar{\mv{y}}|\mv{\phi}) )}{\partial \mv{\phi}}\bigg( \frac{\partial{\rm{ln}}( f(\bar{\mv{y}}|\mv{\phi}) )}{\partial \mv{\phi}}\bigg)^{H}\bigg]\bigg(\frac{\partial {\mv{\phi}}}{\partial \mv{\omega}}\bigg)\bigg]
		%\nonumber\\
		=&\mathbb{E}_{\mv{u}}\bigg[\bigg(\frac{\partial {\mv{\phi}}}{\partial \mv{\omega}}\bigg)^H \mv{J}(\mv{r})\bigg(\frac{\partial {\mv{\phi}}}{\partial \mv{\omega}}\bigg)\bigg], \label{F_Dofomega}
	\end{align}
	where $\mv{J}(\mv{r})\overset{\Delta}{=}\mathbb{E}_{\bar{\mv{y}},\mv{\alpha}}\bigg[\frac{\partial{\rm{ln}}( f(\bar{\mv{y}}|\mv{\phi}) )}{\partial \mv{\phi}}\bigg( \frac{\partial{\rm{ln}}( f(\bar{\mv{y}}|\mv{\phi}) )}{\partial \mv{\phi}}\bigg)^{H}\bigg]$. 
	In (\ref{F_Dofomega}), ${\rm{ln}}(f(\bar{\mv{y}}|\mv{\phi}))=-N{\rm{ln}} (\pi\sigma_z^2)-\frac{1}{\sigma_z^2}\sum_{m=1}^{M}\big(\|\mv{y}_m\|^2+{\hat{p}_m}|\beta_{m}|^2\|{\rm{diag}}(\mv{s}_{m})\mv{g}_m\|^2-2\mathfrak{Re}\{\sqrt{\hat{p}_m}\beta_m^*\mv{g}_m^H{\rm{diag}}(\mv{s}_{m}^H)\mv{y}_m\}\big)$. Moreover, the Jacobian matrix $\frac{\partial {\mv{\phi}}}{\partial \mv{\omega}}\in \mathbb{R}^{3M\times (3+2M)}$ is given by $\frac{\partial {\mv{\phi}}}{\partial \mv{\omega}}=\begin{bmatrix}
		\frac{\partial \mv{r}}{\partial \mv{u}} & \mv{0}\\ 
		\mv{0} & \mv{I}_{2M\times2M}
	\end{bmatrix}$, where $\frac{\partial \mv{r}}{\partial \mv{u}}\in \mathbb{R}^{M\times 3}$ is given by 
	\begin{equation}
		\frac{\partial \mv{r}}{\partial \mv{u}}\! \!=\!\!\begin{bmatrix}
			\frac{\partial r_1}{\partial x} &\frac{\partial r_1}{\partial y}&\frac{\partial r_1}{\partial z} \\ 
			\frac{\partial r_2}{\partial x} &\frac{\partial r_2}{\partial y}&\frac{\partial r_2}{\partial z} \\ 
			\dots&\dots&\dots \\ 
			\frac{\partial r_M}{\partial x} &\frac{\partial r_M}{\partial y}&\frac{\partial r_M}{\partial z}
		\end{bmatrix}\!\!=\!\!\begin{bmatrix}
			\frac{x-\hat{x}_1}{r_1} &\frac{y-\hat{y}_1}{r_1}&\frac{z-\hat{z}_1}{r_1} \\ 
			\frac{x-\hat{x}_2}{r_2} &\frac{y-\hat{y}_2}{r_2}&\frac{z-\hat{z}_2}{r_2} \\ 
			\dots&\dots&\dots \\ 
			\frac{x-\hat{x}_M}{r_M}&\frac{y-\hat{y}_M}{r_M}&\frac{z-\hat{z}_M}{r_M}
		\end{bmatrix}.
	\end{equation}
	Then, we introduce the following proposition to further derive $\mv{J}(\mv{r})$, which serves as a key component in $\mv{F}_{\rm{O}}$.
	\begin{proposition}\label{J_B_Defination}
		$\mv{J}(\mv{r})$ is a diagonal matrix given by
		\begin{align}
			\mv{J}(\mv{r})=\begin{bmatrix}
				{\mv{U}}(\mv{r}) & \mv{0}\\ 
				\mv{0}& \mv{W}(\mv{r})
			\end{bmatrix},
		\end{align}
		where ${\mv{U}}(\mv{r}) ={\rm{diag}}\{\xi_1, \dots, \xi_M\}$ with $\xi_m\!\!=\! \!\! \frac{8P\beta_0^2\sigma_{\alpha}^2}{\sigma_z^2r_{m}^6}+\frac{32\pi^2P\beta_0^2\lambda_{m}\sigma_{\alpha}^2}{c^2\sigma_z^2r_{m}^4}$ and $\lambda_m =\sum_{n=1}^{|\mathcal{N}_m|}\mathcal{N}_m(n)^2\Delta_f^2$ for $m\in\mathcal{M}$; $\mv{W}(\mv{r})$ is a block diagonal matrix given by $\mv{W}(\mv{r}) = [\bar{\mv{W}}(\mv{r}), \mv{0}; \mv{0}, \bar{\mv{W}}(\mv{r})]$ with $\bar{\mv{W}}(\mv{r}) = {\rm{diag}}\{\frac{2P\beta_0^2}{\sigma_z^2r_1^4}, \dots, \frac{2P\beta_0^2}{\sigma_z^2r_M^4}\}$. 
	\end{proposition}
	\begin{IEEEproof}
		Please refer to Appendix  \ref{Proof_J_B_Defination}.
	\end{IEEEproof}
	Based on the above, $\mv{F}_{\rm{O}}$ can be expressed as
	\begin{align}
		\mv{F}_{\rm{O}} =& \mathbb{E}_{\mv{u}} \begin{bmatrix}
			\left(\frac{\partial \mv{r}}{\partial \mv{u}}\right)^{H}{\mv{U}}(\mv{r})\left(\frac{\partial \mv{r}}{\partial \mv{u}}\right) & \mv{0}\\ 
			\mv{0} &  \mv{W}(\mv{r})
		\end{bmatrix}\nonumber\\
		=&\begin{bmatrix}
			\mathbb{E}_{\mv{u}}\left[\left(\frac{\partial \mv{r}}{\partial \mv{u}}\right)^H{\mv{U}}(\mv{r})\left(\frac{\partial \mv{r}}{\partial \mv{u}}\right)\right] & \mv{0}\\ 
			\mv{0} &  \mathbb{E}_{\mv{u}}\left[\mv{W}(\mv{r})\right]
		\end{bmatrix}.
	\end{align}
	Specifically, the top-left sub-matrix in $\mv{F}_{\mathrm{O}}$ is associated with the estimation of $\mv{u}$. It is further expressed as
	\begin{align}
		&\mv{F}_{\mathrm{O}}^{\mv{u}}\overset{\Delta}{=}[\mv{F}_{\rm{O}}]_{1:3, 1:3}\nonumber\\
		=&\int\int\int\sum_{k=1}^K\bar{p}_{X,Y,Z}(x,y,z)\sum_{m=1}^M{{\xi}_m}\left(\frac{\partial \mv{r}}{\partial \mv{u}}\right)^{H}\left(\frac{\partial \mv{r}}{\partial \mv{u}}\right)dxdydz\nonumber
	\end{align}
	\begin{align}
		\approx&\sum_{k=1}^Kp_k \sum_{m=1}^M\bar{\xi}_{m,k}\times\nonumber
		\end{align}
	\begin{align}
		&\!\!\!\!\!\begin{bmatrix}
			{(\hat{x}_m\!\!-\!\!x_k)^2}\!\! & \!\!{(\hat{x}_m\!\!-\!\!x_k)(\hat{y}_m\!\!-\!\!y_k)}\!\!&\!\! {(\hat{x}_m\!\!-\!\!x_k)(\hat{z}_m\!\!-\!\!z_k)}\\ 
			{(\hat{x}_m\!\!-\!\!x_k)(\hat{y}_m\!\!-\!\!y_k)}\! \!&\!\! {(\hat{y}_m\!\!-\!\!y_k)^2}\!\! &\! \!{(\hat{y}_m\!\!-\!\!y_k)(\hat{z}_m\!\!-\!\!z_k)}\\ 
			{(\hat{x}_m\!\!-\!\!x_k)(\hat{z}_m\!\!-\!\!z_k)} \!\!&\!\!{(\hat{y}_m\!\!-\!\!y_k)(\hat{z}_m\!\!-\!\!z_k)} \!\!&\! \!{(\hat{z}_m\!\!-\!\!z_k)^2} 
		\end{bmatrix}\nonumber\\
		\overset{\Delta}{=}&\tilde{\mv{F}}_{\mathrm{O}}^{\mv{u}},\label{F_O_u}
	\end{align}
	where $\bar{\xi}_{m,k} = \frac{8P\beta_0^2\sigma_{\alpha}^2}{\sigma_z^2r_{m,k}^8}+\frac{32\pi^2P\beta_0^2\lambda_{m}\sigma_{\alpha}^2}{c^2\sigma_z^2r_{m,k}^6},\forall m\in\mathcal{M}$ with $r_{m,k}= \sqrt{(\hat{x}_m-x_k)^2+(\hat{y}_m-y_k)^2+(\hat{z}_m-z_k)^2},\forall k\in\mathcal{K}$. Note that the approximation in (\ref{F_O_u}) holds since the probability density under the considered Gaussian mixture PDF $\bar{p}_{X,Y,Z}(x,y,z)$ with a small variance $\sigma^2$ concentrates almost all probability density over $K$ discrete locations $(x_k,y_k,z_k)$'s.
	\addtolength{\topmargin}{0.04in}
	\subsection{Derivation of $\mv{F}_{\mathrm{P}}$}
	Secondly, $\mv{F}_{\rm{P}}= \mathbb{E}_{\mv{\omega}}\left[\frac{\partial {\rm{ln}}p_{\mv{w}}(\mv{\omega})}{\partial \mv{\omega}}\left(\frac{\partial {\rm{ln}}p_{\mv{w}}(\mv{\omega})}{\partial \mv{\omega}}\right)^H\right]\in\mathbb{R}^{(3+2M)\times (3+2M)}$ represents the PFIM extracted from the prior distribution information in $p_{\mv{w}}(\mv{\omega})$. Since all parameters in $\mv{\omega}$ are independent of each other, $\mv{F}_{\mathrm{P}}$ is a block diagonal matrix, which can be expressed as
	\vspace{-0.5mm}\begin{align}
		\mv{F}_{\rm{P}} = \begin{bmatrix}
			\mv{F}_{\rm{P}}^{\mv{u}} & \mv{0}\\ 
			\mv{0} & \mv{F}_{\rm{P}}^{\mv{\alpha}}
		\end{bmatrix}.\vspace{-1mm}
	\end{align}
	The top-left sub-matrix is associated with the prior information for the target's 3D location $\mv{u}$, and is given by 
	\begin{align}
&\mv F_{\rm P}^{\mv u}
= \frac{1}{\sigma^2}\mv I_3
-\frac{1}{2\sigma^4}\sum_{k=1}^{K}\sum_{n=1}^{K}
\iiint \frac{f_k(x,y,z)\,f_n(x,y,z)}{\sum_{\ell=1}^{K} f_\ell(x,y,z)}\times\nonumber\\
&(\mv{u}_k-\mv{u}_n)(\mv{u}_k-\mv{u}_n)^T dx dy dz,\label{F_P_u}
\end{align}
	where $f_k(x,y,z)=\frac{p_k}{(2\pi\sigma^2)^{\frac{3}{2}}}e^{-\frac{(x-x_k)^2+(y-y_k)^2+(z-z_k)^2}{2\sigma^2}}$, and $\mv{u}_k = [x_k,y_k,z_k]^T$, $\forall k\in\mathcal{K}$.
	\subsection{Derivation of Overall PFIM and PCRB}
	The overall PFIM $\mv{F}$ can be approximated as a block-diagonal matrix, given by  
	\begin{align}
		\mv{F}=\begin{bmatrix}
			\mv{F}_{\mathrm{O}}^{\mv{u}}+\mv{F}_{\mathrm{P}}^{\mv{u}}& \mv{0}\\ 
			\mv{0} & \mathbb{E}_{\mv{u}}\left[\mv{W}(\mv{r})\right]+\mv{F}_{\mathrm{P}}^{\mv{\alpha}}
		\end{bmatrix}.
	\end{align}
	The MSE matrix for estimating $\mv{\omega}$, denoted by $\mv{E}$, is lower bounded by the inverse of the overall PFIM $\mv{F}$, i.e., $\mv{E}\succeq \mv{F}^{-1}$. The PCRB for the MSE in estimating the target's 3D location $\mv{u}$ can be further expressed as
	\begin{align}\label{PCBR_value1}
		{\rm{PCRB}}_{\mv{u}} =&{\rm{tr}}([\mv{F}^{-1}]_{1:3,1:3})={\rm{tr}}((\mv{F}_{\mathrm{O}}^{\mv{u}}+\mv{F}_{\mathrm{P}}^{\mv{u}})^{-1})\nonumber\\
		\approx &{\rm{tr}}((\tilde{\mv{F}}_{\mathrm{O}}^{\mv{u}}+\mv{F}_{\mathrm{P}}^{\mv{u}})^{-1}).
	\end{align}
	
	Note that in ${\rm{PCRB}}_{\mv{u}}$, $\tilde{\mv{F}}_{\mathrm{O}}^{\mv{u}}$ is dependent on the BSs' locations and the prior distribution information of the target's 3D location, while $\mv{F}_{\mathrm{P}}^{\mv{u}}$ is a constant matrix determined by the prior distribution information. In the following, we aim to design the BS placement to minimize the PCRB for the MSE in estimating the target's unknown and random 3D location.
	
	\section{Problem Formulation}
	We aim to optimize the horizontal locations of $M$ BSs in the networked sensing system denoted by $\{\hat{x}_m,\hat{y}_m\}_{m=1}^M$ to minimize the PCRB of the MSE in estimating the target's 3D location $\mv{u}$ when its prior distribution information is available for exploitation. The problem is formulated as
	\begin{align}
		{\mbox{(P1)}}\ \underset{\{\hat{x}_m\}_{m=1}^M, \{\hat{y}_m\}_{m=1}^M}{\min} &{\rm{tr}}((\tilde{\mv{F}}_{\mathrm{O}}^{\mv{u}}+\mv{F}_{\mathrm{P}}^{\mv{u}})^{-1}).
	\end{align}
	
	Note that (P1) is a non-convex optimization problem where the objective function involves the trace of inverse of a matrix sum, with each matrix comprising elements that are complex functions of high-order terms with respect to the optimization variables, as shown in (\ref{F_O_u}). Therefore, solving (P1) is highly challenging. In the following, we tackle this problem via equivalent transformations and the inner approximation technique.
	
	\section{Proposed Solution to (P1)}
	\subsection{Equivalent Transformations of (P1)}
	Firstly, we define a set of auxiliary vectors ${\mv{a}}_{m,k} = [\hat{x}_m - x_{k}, \hat{y}_m - y_{k}, \hat{z}_{m}-z_{k}]^T\in \mathbb{R}^{3\times 1}, \forall m\in \mathcal{M}, \forall k\in \mathcal{K}$ to represent the relative location of each BS with respect to each target's possible location. Note that each $[\mv{a}_{m,k}]_{3}=\hat{z}_{m}-z_{k}$ represents the relative height and is a constant. Problem (P1) can be then equivalently transformed into
	\begin{align}
		{\mbox{(P1-eqv-I)}}&\quad \underset{\{\mv{a}_{m, k}\}}{\min}\\
		&\!\!\!\!\!\!\!\!\!\!\!\!\!\!\!\!\!{\rm{tr}}\bigg(\!\bigg(\!\sum_{k=1}^Kp_k\!\sum_{m=1}^M\!\bigg(\!\frac{w\mv{a}_{m,k}\mv{a}_{m,k}^T}{(\mv{a}_{m,k}^T\mv{a}_{m,k})^4}\!\!+\!\!\frac{v_m\mv{a}_{m,k}\mv{a}_{m,k}^T}{(\mv{a}_{m,k}^T\mv{a}_{m,k})^3}\!\bigg)\!\!+\!\!\mv{F}_{\mathrm{P}}^{\mv{u}}\!\bigg)^{-1}\bigg)\nonumber\\
		{\rm{s.t.}} \ \ & \mv{a}_{m,k} - \mv{a}_{m, 1} = [x_1 - x_k, y_1 - y_k, z_1-z_k]^T, \nonumber \\ 
		&\qquad \qquad \  \forall m\in \mathcal{M}, \ \forall k\in\mathcal{K}, k\neq 1 \label{P2_constraint1}\\
		& [\mv{a}_{m,k}]_3 = \hat{z}_{m}-z_k, \ \forall m\in\mathcal{M}, \ \forall k\in\mathcal{K},\label{P2_constraint2}
	\end{align}
	where $w=\frac{8P\beta_0^2\sigma_{\alpha}^2}{\sigma_{z}^2}$, $v_m=\frac{32\pi^2P\beta_0^2\lambda_m\sigma_{\alpha}^2}{c^2\sigma_{z}^2},\forall m\in\mathcal{M}$. Note that the constraints in (\ref{P2_constraint1}) preserve the geometric relationships among all BSs and possible target locations. Then, by introducing auxiliary variables $\{b_{m,k}\}$ and $\{s_{m,k}\}$, we have the following proposition to further simplify (P1-eqv-I).
	\begin{proposition}\label{prop_eqv}
		Problem (P1-eqv-I) and Problem (P1) are equivalent to the following problem:
		\begin{align}\label{P3of1}
			{\mbox{(P1-eqv-II)}} &\min_{\{\mv{a}_{m, k}\},\{b_{m,k}\},\{s_{m,k}\}:(\ref{P2_constraint1}), (\ref{P2_constraint2})}\\ &\!\!\!\!\!\!\!\!\!\!\!\!\!\!\!\!\!\!\!\!\!\!\!\!\!\!\!{\rm{tr}}\bigg(\bigg(\sum_{k=1}^Kp_k\sum_{m=1}^M\bigg(\frac{w\mv{a}_{m,k}\mv{a}_{m,k}^T}{s_{m,k}}\!+\!\frac{v_m\mv{a}_{m,k}\mv{a}_{m,k}^T}{b_{m,k}}\bigg)\!+\!\mv{F}_{\mathrm{P}}^{\mv{u}}\bigg)^{-1}\bigg)\nonumber\\
			\ {\rm{s.t.}}\  \ 
			&b_{m,k}^{\frac{1}{3}} \geq\mv{a}_{m,k}^T \mv{a}_{m,k},\quad \forall m \in \mathcal{M}, \ \forall k \in \mathcal{K}\label{P3_constraint1}\\
			&s_{m,k}^{\frac{1}{4}} \geq\mv{a}_{m,k}^T \mv{a}_{m,k},\quad \forall  m \in \mathcal{M}, \ \forall k \in \mathcal{K}.\label{P3_constraint2}
		\end{align}
	\end{proposition}
	\begin{IEEEproof}
		Given any feasible solution $\{\mv{a}_{m,k}\}$ for (P1-eqv-I), $b_{m,k} = ({\mv{a}_{m,k}}^T\mv{a}_{m,k})^3, s_{m,k} = ({\mv{a}_{m,k}}^T\mv{a}_{m,k})^4, \forall m,k$ is a feasible solution to (P1-eqv-II) with the same objective value. Thus, the optimal value of (P1-eqv-II) is no larger than that of (P1-eqv-I). Moreover, since the objective value of (P1-eqv-II) can be shown to be monotonically increasing with each $b_{m,k}$ or $s_{m,k}$, the constraints in (\ref{P3_constraint1}) and (\ref{P3_constraint2}) must hold with equality at the optimum, i.e., ${b_{m,k}^{\star}}^{\frac{1}{3}}={\mv{a}_{m,k}^{\star}}^T \mv{a}_{m,k}^{\star}, {s_{m,k}^{\star}}^{\frac{1}{4}}={\mv{a}_{m,k}^{\star}}^T \mv{a}_{m,k}^{\star}, \forall m,k$, ensuring that the optimal value of (P1-eqv-II) is no smaller than that of (P1-eqv-I).  Therefore, the optimal values of (P1-eqv-II), (P1-eqv-I), and consequently (P1) are equal, which completes the proof of Proposition \ref{prop_eqv}. 
	\end{IEEEproof}
	
	Then, by applying the Schur complement technique, Problem (P1-eqv-II) can be equivalently transformed into 
	\begin{align}
		&{\mbox{(P1-eqv-III)}}\ 
		\min_{\mv{T}\!,\{\mv{a}_{m\!, k}\},\{b_{m,k}\},\{s_{m,k}\}}
		{\rm{tr}}(\mv{T})\\
		&{\rm{s.t.}}\ \begin{bmatrix}
			\!\sum_{k=1}^K \! p_k\!\sum_{m=1}^M\!\!\bigg(\!\frac{w\mv{a}_{m,k}\mv{a}_{m,k}^T}{s_{m,k}}\!+\!\frac{v_m\mv{a}_{m,k}\mv{a}_{m,k}^T}{b_{m,k}}\bigg)\!+\!\mv{F}_{\mathrm{P}}^{\mv{u}}\!\!& \!\!\mv{I}\\
			\mv{I}  &  \mv{T}
		\end{bmatrix}\nonumber\\
		&\qquad \succeq  \mv{0}\label{P3_1_constraint1}\\
		&\qquad	(\ref{P2_constraint1}), (\ref{P2_constraint2}), (\ref{P3_constraint1}),(\ref{P3_constraint2}).
	\end{align}
	
	In the following, we will deal with (P1-eqv-III), which is equivalent to the original problem (P1).
	\begin{figure*}
		\begin{align}\label{ineq}
			\begin{bmatrix}
				\!\sum_{k=1}^K \! p_k\!\sum_{m=1}^M\!\!\bigg(\!\frac{w\mv{a}_{m,k}\mv{a}_{m,k}^T}{s_{m,k}}\!+\!\frac{v_m\mv{a}_{m,k}\mv{a}_{m,k}^T}{b_{m,k}}\bigg)\!+\!\mv{F}_{\mathrm{P}}^{\mv{u}}\!\!& \!\!\mv{I}\\
				\mv{I}  &  \mv{T}
			\end{bmatrix}\!\succeq\! \begin{bmatrix} \sum_{k=1}^K p_k \sum_{m=1}^M\hat{\mv{M}}\!(\mv{a}_{m,k}, \!b_{m,k}, \!s_{m,k}|\mv{a}_{m,k}^{(i)}, b_{m,k}^{(i)}, s_{m,k}^{(i)})\! +\! \mv{F}_{\mathrm{P}}^{\mv{u}}\!\! & \!\!\!\mv{I}\\
				\mv{I} &\! \!\!\mv{T}
			\end{bmatrix}.
		\end{align}
		\vspace{-7mm}
	\end{figure*}
	\vspace{-2mm}
	\subsection{Iterative Inner Approximation for (P1-eqv-III)}
The key difficulty in (P1-eqv-III) lies in the non-convex constraint in (\ref{P3_1_constraint1}). To overcome this issue, we employ the \emph{iterative inner approximation method} \cite{marks1978general}, which successively approximates the original non-convex feasible set by a sequence of convex ones. Specifically, consider a non-convex matrix inequality constraint $\mv{h}(\mv{t})\preceq \mv{0}$, where $\mv{t}$ denotes the optimization variable. Let $\bar{\mv{h}}(\mv{t}|\mv{t}^{(i)})$ denote an inner approximation of $\mv{h}(\mv{t})$ at the $i$-th iteration, where $\mv{t}^{(i)}$ is the solution from the $i$-th iteration. The inner approximation iterations are guaranteed to converge to a KKT point of the original problem as long as $\bar{\mv{h}}(\mv{t}|\mv{t}^{(i)})$ satisfies the following conditions \cite{marks1978general}: 
	\begin{itemize}
		\item $\bar{\mv{h}}(\mv{t}^{(i)}|\mv{t}^{(i)})=\mv{h}(\mv{t}^{(i)})$
		\item $\nabla_{\mv{t}}\bar{\mv{h}}(\mv{t}|\mv{t}^{(i)})|_{\mv{t} = \mv{t}^{(i)}} = \nabla_{\mv{t}}\mv{h}(\mv{t})|_{\mv{t} =\mv{t}^{(i)}}$
		\item $\bar{\mv{h}}(\mv{t}|\mv{t}^{(i)})\succeq \mv{h}(\mv{t})$.
	\end{itemize}
	
	To construct a convex inner approximation of the constraint in (\ref{P3_1_constraint1}), we introduce the following lemma. 
	\begin{lemma}\label{lemma1} For $\mv{p}, \mv{p}_{i}\in\mathbb{R}^{N\times1}$ and $q, q_{i}\in\mathbb{R}^{++}$, the following inequality holds: 
		\begin{align}
			q^{-1}\mv{p}\mv{p}^{T}\succeq q_{i}^{-1}\mv{p}{\mv{p}_{i}}\!^{T} + q_{i}^{-1}\mv{p}_{i}\mv{p}^{T}- q_{i}^{-2}q\mv{p}_{i}{\mv{p}_{i}}\!^{T}, 
		\end{align}
		where equality is achieved if and only if $\mv{p}=\mv{p}_{i}$ and $q=q_{i}$.
	\end{lemma}
	\begin{IEEEproof}
		Please refer to Appendix \ref{Proof_Lemma_Equation}.
	\end{IEEEproof}
	
	Inspired by the above, we construct the following surrogate functions for the inner approximation:
	\begin{align}
		&\mv{M}(\mv{a}_{m,k}, b_{m,k}|\mv{a}_{m,k}^{(i)}, b_{m,k}^{(i)})\nonumber\\
		=&\frac{\mv{a}_{m,k} {\mv{a}_{m,k}^{(i)}}\!^T}{{b}_{m,k}^{(i)}}\!+\!\frac{\mv{a}_{m,k}^{(i)} {\mv{a}_{m,k}}^T}{{b}_{m,k}^{(i)}}-\frac{b_{m,k}\mv{a}_{m,k}^{(i)} {\mv{a}_{m,k}^{(i)}}\!^T}{{{b}_{m,k}^{(i)}}^2},\\
		&\mv{M}(\mv{a}_{m,k}, s_{m,k}|\mv{a}_{m,k}^{(i)}, s_{m,k}^{(i)})\nonumber\\
		=&\frac{\mv{a}_{m,k} {\mv{a}_{m,k}^{(i)}}\!^T}{{s}_{m,k}^{(i)}}\!+\!\frac{\mv{a}_{m,k}^{(i)} {\mv{a}_{m,k}}^T}{{s}_{m,k}^{(i)}}-\frac{s_{m,k}\mv{a}_{m,k}^{(i)} {\mv{a}_{m,k}^{(i)}}\!^T}{{{s}_{m,k}^{(i)}}\!^2}.
	\end{align}
	Note that $\mv{M}(\mv{a}^{(i)}_{m,k}, b^{(i)}_{m,k}|\mv{a}_{m,k}^{(i)}, b_{m,k}^{(i)})\!\!=\!\!\frac{\mv{a}_{m,k}^{(i)}{\mv{a}_{m,k}^{(i)}}\!^T}{{b}_{m,k}^{(i)}}$ and $\mv{M}(\mv{a}^{(i)}_{m,k}, s^{(i)}_{m,k}|\mv{a}_{m,k}^{(i)}, s_{m,k}^{(i)})\!\!=\!\!\frac{\mv{a}_{m,k}^{(i)}{\mv{a}_{m,k}^{(i)}}\!^T}{{s}_{m,k}^{(i)}}$. Moreover, we have
	\begin{align}
		&\nabla_{\mv{t}_{m,k}}\frac{\mv{a}_{m,k} \mv{a}_{m,k}^T}{{b}_{m,k}}\big|_{\mv{t}_{m,k}=\mv{t}_{m,k}^{(i)}}\nonumber\\
		=&\nabla_{\mv{t}_{m,k}}\mv{M}(\mv{a}_{m,k},{b}_{m,k}|\mv{a}_{m,k}^{(i)},{b}_{m,k}^{(i)})\big|_{\mv{t}_{m,k}=\mv{t}_{m,k}^{(i)}},
	\end{align}
	\begin{align}
		&\nabla_{\mv{z}_{m,k}}\frac{\mv{a}_{m,k} \mv{a}_{m,k}^T}{{s}_{m,k}}\big|_{\mv{z}_{m,k}=\mv{z}_{m,k}^{(i)}}\nonumber\\
		=&\nabla_{\mv{z}_{m,k}}\mv{M}(\mv{a}_{m,k},{s}_{m,k}|\mv{a}_{m,k}^{(i)},{s}_{m,k}^{(i)})\big|_{\mv{z}_{m,k}=\mv{z}_{m,k}^{(i)}},
	\end{align}
	with $\mv{t}_{m,k} = [\mv{a}_{m,k}^T, b_{m,k}]^T$, $\mv{t}_{m,k}^{(i)} = [{\mv{a}_{m,k}^{(i)}}^T, b_{m,k}^{(i)}]^T$, $\mv{z}_{m,k} = [\mv{a}_{m,k}^T, s_{m,k}]^T$, and $\mv{z}_{m,k}^{(i)} = [{\mv{a}_{m,k}^{(i)}}^T, s_{m,k}^{(i)}]^T$. By noting that $b_{m,k}>0, s_{m,k}>0, \forall m,k$, we further have the following condition according to Lemma \ref{lemma1}:
	\begin{align}
		&\frac{v_m\mv{a}_{m,k} \mv{a}_{m,k}^T}{{b}_{m,k}}\!+\!\frac{w\mv{a}_{m,k} \mv{a}_{m,k}^T}{{s}_{m,k}}
		\!\succeq \!
		v_m\mv{M}(\mv{a}_{m,k}, b_{m,k}|\mv{a}_{m,k}^{(i)}, b_{m,k}^{(i)})\nonumber\\
		&+w\mv{M}(\mv{a}_{m,k}, s_{m,k}|\mv{a}_{m,k}^{(i)}, s_{m,k}^{(i)})\nonumber\\
		&\overset{\Delta}{=}\hat{\mv{M}}(\mv{a}_{m,k}, b_{m,k}, s_{m,k}|\mv{a}_{m,k}^{(i)}, b_{m,k}^{(i)}, s_{m,k}^{(i)}).
	\end{align}
	Therefore, we have the inequality in (\ref{ineq}). This shows that the right-hand side (RHS) of (\ref{ineq}) satisfies the three conditions above, thus being a suitable convex inner approximation of the left-hand side (LHS) of (\ref{P3_1_constraint1}).
	
	By replacing the LHS of (\ref{P3_1_constraint1}) with the RHS of (\ref{ineq}), we formulate the following problem, which is the inner approximated version of the original problem (P1-eqv-III) in the $(i+1)$-th iteration based on local point $(\mv{a}_{m,k}^{(i)}, b_{m,k}^{(i)}, s_{m,k}^{(i)})$:
	\begin{align}
		&{\mbox{(P1-IA)}}\quad \min_{\mv{T},\{\mv{a}_{m, k}\},\{b_{m,k}\},\{s_{m,k}\}} \quad  {\rm{tr}}(\mv{T})\\
		&{\rm{s.t.}} \nonumber\\
		&\!\!\begin{bmatrix}\!\! \sum_{k=1}^K\! p_k\! \sum_{m=1}^M\!\!\hat{\mv{M}}\!(\!\mv{a}_{m,k}, \!b_{m,k}, \!s_{m,k}|\mv{a}_{m,k}^{(i)},\! b_{m,k}^{(i)},\! s_{m,k}^{(i)}\!)\!\! +\!\! \mv{F}_{\mathrm{P}}^{\mv{u}} \!\!& \!\!\!\mv{I}\\
			\mv{I} \!\!&\! \!\!\mv{T}
		\end{bmatrix}\nonumber\\
		&\succeq\!\mv{0} \label{P3_IIconstraint}\\
		&(\ref{P2_constraint1}), (\ref{P2_constraint2}), (\ref{P3_constraint1}),(\ref{P3_constraint2}).
	\end{align}
	Problem (P1-IA) is a convex optimization problem which can be solved via the interior-point method \cite{bibConvexOpt} or CVX \cite{cvx}.
	\vspace{-2mm}
	\subsection{Overall Algorithm, Quality Guarantee, and Complexity}	\vspace{-2mm}
	In Algorithm \ref{algorithm 1}, we summarize the proposed iterative inner approximation based algorithm for the BS placement optimization (P1). Specifically, starting from a set of initial solutions of $(\{\mv{a}_{m,k}\},\{b_{m,k}\},\{s_{m,k}\})$ denoted by $(\{\mv{a}_{m,k}^{(0)}\},\{b_{m,k}^{(0)}\},\{s_{m,k}^{(0)}\})$, the algorithm iteratively solves (P1-IA) and updates $(\{\mv{a}_{m,k}^{(i)}\},\{b_{m,k}^{(i)}\},\{s_{m,k}^{(i)}\})$ as the optimal solution of $(\{\mv{a}_{m,k}\},\{b_{m,k}\},\{s_{m,k}\})$ in the $i$-th iteration. The algorithm is guaranteed to converge to a KKT point of (P1-eqv-III) \cite{marks1978general}. Moreover, we have the following proposition to unveil the solution quality for the original problem (P1-eqv-I). 
	\begin{proposition}\label{KKT_point}
		The KKT point $\{\mv{a}_{m,k}\}$ for (P1-eqv-III) obtained via Algorithm \ref{algorithm 1} is also a KKT point for (P1-eqv-I).
	\end{proposition}
	\begin{IEEEproof}
		Please refer to Appendix \ref{Proof_KKT_point}.
	\end{IEEEproof}
	Furthermore, since the variables transformation from (P1) to (P1-eqv-I) is both affine and bijective, there exists a one-to-one correspondence between their KKT points. Consequently, a KKT point of the original problem (P1) can be obtained based on the KKT point $\{\mv{a}_{m,k}\}$ for (P1-eqv-III) obtained via Algorithm \ref{algorithm 1} according to the relationship between $\{\mv{a}_{m,k}\}$ and $\{\hat{x}_m,\hat{y}_m\}_{m=1}^M$ specified at the beginning of Section V.
	
In (P1-IA), there are $(K-1)M + 3KM + 1$ constraints, $5MK$ variables, and one $3 \times 3$ matrix variable. The complexity for solving (P1-IA) using the interior-point method is given by $\mathcal{O}(((5MK+9)^3 + ((K-1)M + 3KM + 1)(5MK+9)^2)\sqrt{5MK+9}) \propto \mathcal{O}(M^{3.5} K^{4.5})$. 
Let $N_{\rm{I}}$ represent the total number of iterations for the inner approximation method. The overall complexity for Algorithm \ref{algorithm 1} is thus given by $\mathcal{O}(N_{\rm{I}}M^{3.5} K^{4.5})$.

	%Finally, it is worth noting that the proposed algorithm is directly applicable to the case where various practical site constraints are imposed on the BS locations, such as a minimum distance requirement between BSs and/or between BSs and target's possible locations, an 
	
	\begin{algorithm}[t]
		\renewcommand{\algorithmicrequire}{\textbf{Input:}}
		\renewcommand{\algorithmicensure}{\textbf{Output:}}
		\caption{Proposed inner approximation algorithm for (P1)}	\label{alg1}
		\begin{algorithmic}[1]\label{algorithm 1}
			%\REQUIRE $M\!$, \!$K\!$, \!$\!\{p_k\}$, \!$\{\lambda_m\}$, $\{\!(x_k, y_k, z_k)\!\}$, \!$\!\{\hat{z}_m\}$, \!$\!\sigma_{\alpha}^2$, \!$\!\sigma_z^2$, \!$\!P$, \!$\!\sigma^2$.
			\STATE \textbf{Initialization}: Set $i = 0$. Initialize $\{\mv{a}_{m,k}^{(0)}\}$, $\{s_{m,k}^{(0)}\}$,\\
			and $\{b_{m,k}^{(0)}\}$.
			%\STATE Calculate the initial objective value, $g(\{\mv{a}_{m,k}^{(0)}\}, \{s_{m,k}^{(0)}\}, \{b_{m,k}^{(0)}\})$
			\REPEAT
			\STATE Solve (P1-IA) based on $\{\mv{a}_{m,k}^{(i)}, s_{m,k}^{(i)}, b_{m,k}^{(i)}\}$.
			\STATE Update $\{\mv{a}_{m,k}^{(i+1)}, s_{m,k}^{(i+1)}, b_{m,k}^{(i+1)}\}$ as the obtained optimal solutions to (P1-IA).
			%\STATE Update variables: 
			%	$\{\mv{a}_{m,k}^{(i+1)}\} = \{\mv{a}_{m,k}^{\star}\}, \{s_{m,k}^{(i+1)}\} = \{s_{m,k}^{\star}\},$ and $\{b_{m,k}^{(i+1)}\} = \{b_{m,k}^{\star}\}$.
			%	\STATE Compute the new objective value $g(\{\mv{a}_{m,k}^{(i+1)}\}, \{s_{m,k}^{(i+1)}\}, \{b_{m,k}^{(i+1)}\})$.
			%\STATE Check convergence: If
			%	$\frac{|\!g(\{\mv{a}_{m,k}^{(i+1)}\}, \{s_{m,k}^{(i+1)}\}, \{b_{m,k}^{(i+1)}\}\!)\! - \!g(\!\{\mv{a}_{m,k}^{(i)}\}, \{s_{m,k}^{(i)}\}, \{b_{m,k}^{(i)}\}\!)\!|}{g(\{\mv{a}_{m,k}^{(i+1)}\}, \{s_{m,k}^{(i+1)}\}, \{b_{m,k}^{(i+1)}\})}\!\!\leq\!\!\epsilon\!$,
			%	then \textbf{terminate}; otherwise, set $i \leftarrow i+1$ and go to Step 4.
			\UNTIL Convergence.
			\STATE Recover the BS locations $\{(\hat{x}_m, \hat{y}_m)\}$ based on the converged $\{\mv{a}_{m,k}\}$ and $\{(x_k, y_k, z_k)\}$.
			%	\ENSURE $\{(\hat{x}_{m}^{\star},\hat{y}_{m}^{\star})\}$.
		\end{algorithmic} 
		\vspace{-1mm}
	\end{algorithm} 
	\vspace{-3mm}
	\section{Numerical Results}
	\vspace{-2mm}
	In this section, we present numerical results to evaluate the performance of the proposed BS placement scheme. We consider a system with $M=4$ BSs, where all BS have the same height of $\hat{z}_1=\hat{z}_2 = \hat{z}_{3} = \hat{z}_4=20$ (m). We consider $N=2048$  sub-carriers in the OFDM system and $\Delta_f = 30~{\rm{kHz}}$. Moreover, each $m$-th BS is assigned with $|\mathcal{N}_m| = 512$ sub-carriers with $\mathcal{N}_{m} = \{0+m,4+m,\dots, 2044+m\}$.	We consider $K=4$ possible target locations given by $(x_1, y_1, z_1)\!\!=\!\! (0, 0, 13)$, $(x_2, y_2, z_2)\!\!=\!\! (8, 20, 13)$, $(x_3, y_3, z_3) \!=\! (40, 18, 10)$, and $(x_4, y_4, z_4) \!=\! (45, 25, 3)$ (m). The corresponding probabilities are $p_1 \!=\! 0.25$, $p_2 \!=\! 0.2$, $p_3 \!=\! 0.3$, and $p_4 \!=\! 0.25$. In the Gaussian mixture PDF approximation, we set the variance as $\sigma^2\!=\! 10^{-4}$. We further set $P \!=\! 20\ {\rm{dBm}}$, $\sigma_z^2 \!=\! -90\ {\rm{dBm}}$, $\beta_0\!=\!-30$ dB, and $\sigma_\alpha^2 \!=\! 1$. The convergence criterion for Algorithm \ref{algorithm 1} is that the relative decrement in the objective value is no larger than $10^{-7}$.
	
\begin{figure}[t]
		\centering
		\includegraphics[width=6cm]{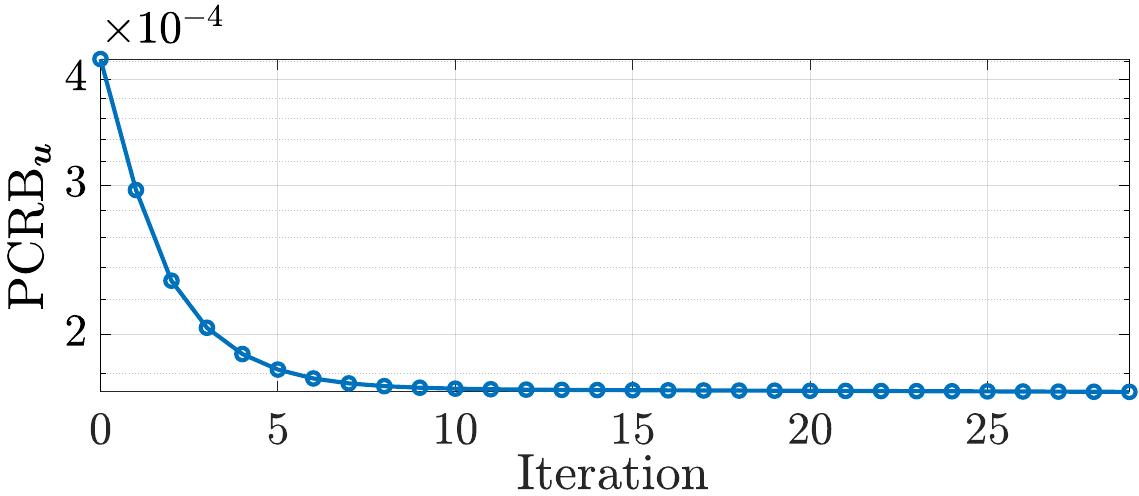}
				\vspace{-2mm}
		\caption{Convergence of Algorithm 1.}
		\label{iterations}
						\vspace{-3mm}
	\end{figure}

	Fig. \ref{iterations} illustrates the convergence behavior of Algorithm \ref{algorithm 1}. It is observed that Algorithm \ref{algorithm 1} converges monotonically and quickly within 30 iterations. Then, we compare our proposed BS placement design with two benchmark schemes:
	\begin{itemize}
		\item {\bf{Benchmark I}}: The BS locations are optimized assuming the target's location is known and fixed at $(40, 18, 10)$ m (the most probable location). The horizontal locations of BSs are then uniformly distributed on a circle of radius $r_0$ centered at the assumed target location \cite{godrich2010target}.
		\item {\bf{Benchmark II}}: The BS placement is determined by sequentially minimizing the sum of distances to subsets of possible target locations that are dynamically selected based on their cumulative probabilities. Specifically, BS $1$'s location is optimized by minimizing its sum distance to all $K$ possible target locations, i.e., $\underset{(\hat{x}_{1},\hat{y}_{1})}{\min}\sum_{k\in\mathcal{K}}\sqrt{(\hat{x}_{1} - x_k)^2+(\hat{y}_{1}-y_k)^2+(\hat{z}_{1}-z_k)^2}$. Then, BS $2$'s location is optimized by minimizing its sum of distances to the $K-1$ possible target locations with the highest combined probabilities; BS $3$'s location is optimized by minimizing its sum of distances with respect to the $K-1$ possible target locations with the second-highest combined probabilities; so on and so forth. In each step, the location optimization for each BS is a generalized Fermat-Weber problem which can be solved via a Weiszfeld-like iteration method \cite{beck2015weiszfeld}.
	\end{itemize}\vspace{-1mm}
	
\begin{figure}[t]
	\centering
	\includegraphics[width=6cm]{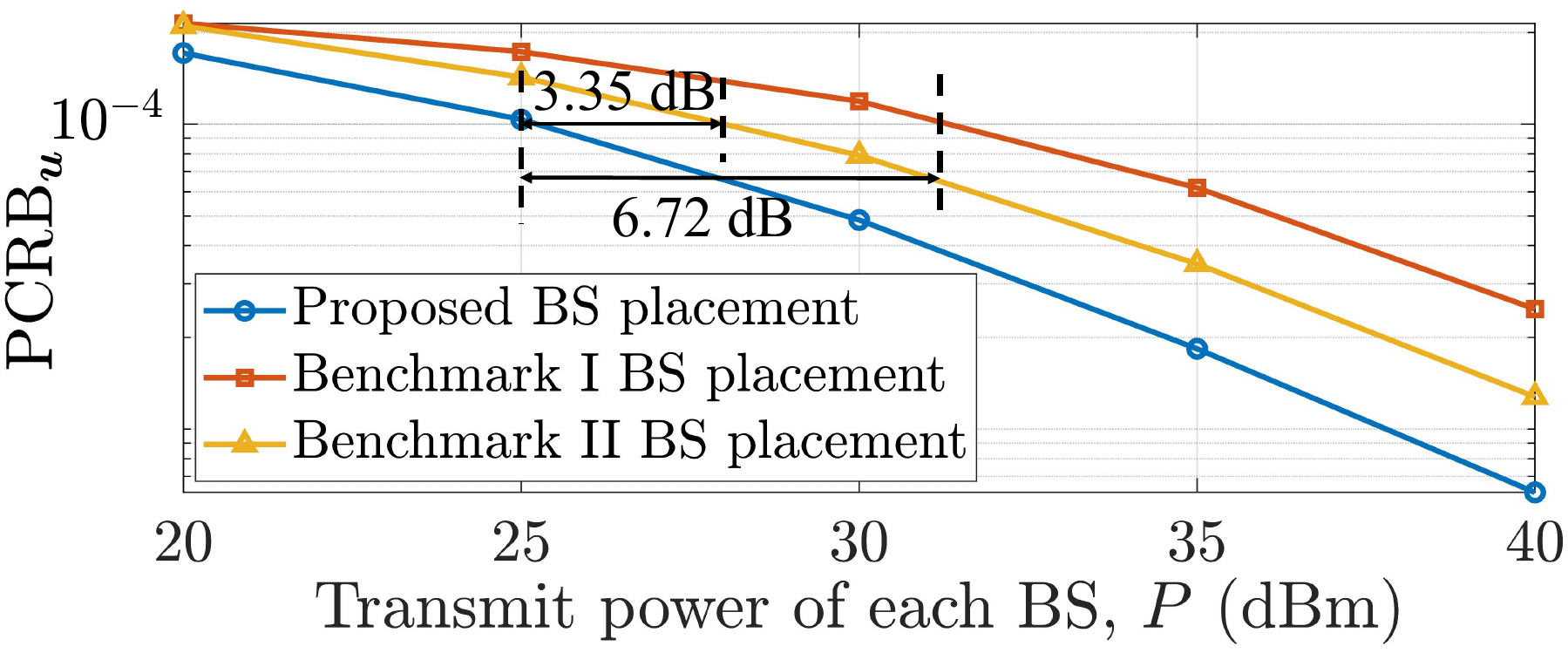}
			\vspace{-2mm}
	\caption{Performance of different BS placement schemes.}
	\vspace{-2mm}
	\label{PCRBvPower}
\end{figure}
	
\begin{figure}[t]
		\centering
		\includegraphics[width=6cm]{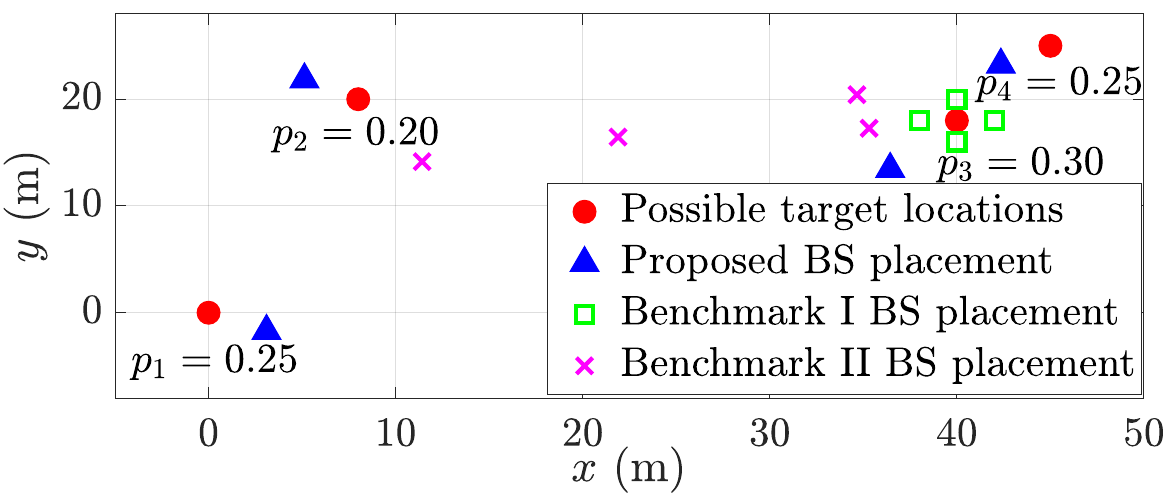}
		\vspace{-2mm}
		\caption{Illustration of different BS placement schemes.}\vspace{-2mm}
		\label{Placement}
	\end{figure}

	Fig. \ref{PCRBvPower} shows the sensing PCRB versus transmit power at each BS for the proposed and benchmark schemes, with $r_0=2$ m for Benchmark I. The corresponding BS placements projected on the horizontal plane are illustrated in Fig. \ref{Placement}. It is observed from Fig. \ref{PCRBvPower} that our proposed scheme outperforms Benchmark II and Benchmark I by $3.35$ dB and $6.72$ dB at a PCRB level of $10^{-4}$, respectively. Furthermore, it is observed from Fig. \ref{Placement} that by only considering the most probable target location, Benchmark I yields a large distance between the BSs and all other possible target locations; while by sequentially designing the location of each BS, Benchmark II distributes BSs approximately uniformly near the center of all possible target locations, leading to large distances between each BS and possible target locations. In contrast, by jointly optimizing all BS locations according to the PCRB closed-form expression derived in Section III, our proposed placement is able to capture the effect of both networked sensing and prior information, thus yielding a balanced distance distribution among each pair of BS and target possible location with higher priority over highly-probable target locations.
	
	\vspace{-1mm}
	\section{Conclusions}\vspace{-1mm}
	This paper studies a multi-BS OFDM-based networked sensing system for estimating the unknown and random 3D location of a target, whose distribution information is available for exploitation. We first characterize the PCRB of the MSE in estimating the target's 3D location as an explicit function of the BS locations and target location distribution in closed form. Based on this, we study the placement optimization of the multiple BSs to minimize the sensing PCRB. Although the problem is non-convex and difficult to solve, we devise an iterative inner approximation based algorithm which is guaranteed to converge to a KKT point of the original problem. Numerical results show that our proposed placement design significantly outperforms various benchmark designs.
	
\vspace{-1mm}
	\appendix
	\subsection{Proof of Proposition \ref{J_B_Defination}}\label{Proof_J_B_Defination}

To derive $\mv{J}(\mv{r}) = \mathbb{E}_{\mv{\alpha}}\left[\mathbb{E}_{\bar{\mv{y}}}\left[\!\frac{\partial{\rm{ln}}\left( \!f\left(\bar{\mv{y}}|\mv{\phi}\right) \!\right)}{\partial \mv{\phi}}\!\!\left(\!\!\frac{\partial{\rm{ln}}\left( \!f\left(\bar{\mv{y}}|\mv{\phi}\right) \!\right)}{\partial \mv{\phi}}\!\right)^{\!H}\right]\!\right]$, we first derive the inner expectation matrix. To this end, we define the inner expectation matrix as a block-structured matrix given by
	\vspace{-1mm}\begin{align}\label{J_phi}
		\hspace{-2mm}\mathbb{E}_{\bar{\mv{y}}}\!\!\left[\!\frac{\partial{\rm{ln}}\left( \!f\left(\bar{\mv{y}}|\mv{\phi}\right) \!\right)}{\partial \mv{\phi}}\!\!\left(\!\!\frac{\partial{\rm{ln}}\left( \!f\left(\bar{\mv{y}}|\mv{\phi}\right) \!\right)}{\partial \mv{\phi}}\!\right)^{\!H}\!\right]\!\!\triangleq\!\!\begin{bmatrix}
			\mv{U}_{\rm{r}}&\mv{B}_{\rm{R}}&\mv{B}_{\rm{I}}\\
			\mv{B}_{\rm{R}}^H&\mv{U}_{\rm{R}}&\mv{V}\\
			\mv{B}_{\rm{I}}^H&\mv{V}^H&\mv{U}_{\rm{I}}
		\end{bmatrix}\!\!.\!\!
	\end{align}
Note that all block matrices are diagonal, with their corresponding diagonal elements given by $\left[\mv{U}_{\rm{r}}\right]_{m,m} =\mathbb{E}_{\bar{\mv{y}}}\left[\frac{\partial f(\bar{\mv{y}}|\mv{\phi})}{\partial{r}_{m}}\left(\frac{\partial f(\bar{\mv{y}}|\mv{\phi})}{\partial{r}_{m}}\right)^H\right]= \frac{8P\beta_0^2|\alpha_{m}|^2}{\sigma_z^2r_{m}^6}+\frac{32\pi^2P\beta_0^2\lambda_{m}|\alpha_{m}|^2}{c^2\sigma_z^2r_{m}^4}$, where $\lambda_m =\sum_{n=1}^{|\mathcal{N}_m|} \mathcal{N}_{m}(n)^2\Delta_f^2$, 
	$\left[\mv{B}_{\rm{R}}\right]_{m, m} =\mathbb{E}_{\bar{\mv{y}}}\left[\frac{\partial f(\bar{\mv{y}}|\mv{\phi})}{\partial r_{m}}\left(\frac{\partial f(\bar{\mv{y}}|\mv{\phi})}{\partial{\alpha}_{m}^{\rm{R}}}\right)^H\right]= \frac{2P\beta_0^2\alpha^{\rm{R}}_{m_1}}{\sigma_z^2r_{m}^5}$,
	$\left[\mv{U}_{\rm{I}}\right]_{m, m}=\left[\mv{U}_{\rm{R}}\right]_{m, m} =\mathbb{E}_{\bar{\mv{y}}}\left[\frac{\partial f(\bar{\mv{y}}|\mv{\phi})}{\partial {\alpha}_{m}^{\rm{R}}}\left(\frac{\partial f(\bar{\mv{y}}|\mv{\phi})}{\partial{\alpha}_{m}^{\rm{R}}}\right)^H\right]= \frac{2P\beta_0^2}{\sigma_z^2r_{m}^4}$,
	$\left[\mv{B}_{\rm{I}}\right]_{m, m}\!=\!\mathbb{E}_{\bar{\mv{y}}}\left[\frac{\partial f(\bar{\mv{y}}|\mv{\phi})}{\partial r_{m}}\left(\frac{\partial f(\bar{\mv{y}}|\mv{\phi})}{\partial{\alpha}_{m}^{\rm{I}}}\right)^H\right] = \frac{2P\beta_0^2\alpha^{\rm{I}}_{m}}{\sigma_z^2r_{m}^5}$, 
	$\left[\mv{V}\right]_{m, m}=\mathbb{E}_{\bar{\mv{y}}}\left[\frac{\partial f(\bar{\mv{y}}|\mv{\phi})}{\partial {\alpha}_{m}^{\rm{R}}}\left(\frac{\partial f(\bar{\mv{y}}|\mv{\phi})}{\partial{\alpha}_{m}^{\rm{I}}}\right)^H\right]=0,\ \forall m\in\mathcal{M}$.
	Then, taking the expectation over $\mv{\alpha}$, we can obtain $\mv{U}({\mv{r}})=\mathbb{E}_{\mv{\alpha}}[\mv{U}_{r}] = {\rm{diag}}\{\xi_1, \dots, \xi_{M}\}$ with $\xi_m = \frac{8P\beta_0^2\sigma_{\alpha}^2}{\sigma_z^2r_{m}^6}+\frac{32\pi^2P\beta_0^2\lambda_{m}\sigma_{\alpha}^2}{c^2\sigma_z^2r_{m}^4},\ \forall m\in\mathcal{M}$; $\bar{\mv{W}}(\mv{r}) = \mathbb{E}_{\mv{\alpha}}[\mv{U}_{\rm{R}}]=\mathbb{E}_{\mv{\alpha}}[\mv{U}_{\rm{I}}]={\rm{diag}}\{\frac{2P\beta_0^2}{\sigma_z^2r_1^4}, \dots, \frac{2P\beta_0^2}{\sigma_z^2r_M^4}\}$; and $\mathbb{E}_{\mv{\alpha}}[\mv{V}]=\mathbb{E}_{\mv{\alpha}}[\mv{B}_{\rm{R}}] =\mathbb{E}_{\mv{\alpha}}[\mv{B}_{\rm{I}}] = \mv{0}$. Proposition \ref{J_B_Defination} is thus proved. 
	\vspace{-2mm}
	\subsection{Proof of Lemma \ref{lemma1}}
	\vspace{-2mm}
	\label{Proof_Lemma_Equation}
	Since $\left(q^{-1}\mv{a} -{q_{i}}^{-1}{\mv{p}_{i}}\!^{-1}\right)\left(q^{-1}\mv{p} - {q_{i}}\!^{-1}{\mv{p}_{i}}\!^{-1}\right)^{T}\succeq\mv{0}$, we have $q^{-2}\mv{p}\mv{p}^{T} - q^{-1}{q_{i}}\!^{-1}\mv{p}{\mv{p}_{i}}\!^{T} - q^{-1}q_{i}\!^{-1}\mv{p}_{i}\mv{p}^{T}+{q_{i}}\!^{-2}\mv{p}_{i}\mv{p}_{i}^T\succeq\mv{0}$. Since $q>0$, we have $q^{-1}\mv{p}\mv{p}^{T}\succeq q_{i}^{-1}\mv{p}\mv{p}_{i}^{T} + q_{i}^{-1}\mv{p}_{i}\mv{a}^{T}- q_{i}^{-2}q\mv{p}_{i}\mv{p}_{i}^{T}$. The proof of Lemma \ref{lemma1} is thus completed.
	
	\subsection{Proof of Proposition \ref{KKT_point}}
	\label{Proof_KKT_point}
	Fixing $\{s_{m,k}\}$, if the KKT point $\{b_{m,k}^{\star}\}$ of (P1-eqv-III) is also the KKT point of (P1-eqv-I), we can prove that the conclusion holds for $\{s_{m,k}\}$. Thus, we focus on the proof on variables $\{b_{m,k}^{\star}\}$ as follows.
	
	First, we analyze the solutions satisfying KKT conditions of (P1-eqv-III) and (P1-eqv-II) using the Lagrangian duality method.
	Since the constraints in (\ref{P2_constraint1}) and (\ref{P2_constraint2}) are identical for two problems (P1-eqv-III) and (P1-eqv-II), we can ignore these shared constraints in the subsequent analysis. 
	
	Let $\{\lambda_{m,k}^{\rm{I}}\}$ and $\{\eta_{m,k}^{\rm{I}}\}$ denote the dual variables associated with the constraints in (\ref{P3_constraint1}) and (\ref{P3_constraint2}) in (P1-eqv-II), respectively, where $\lambda_{m,k}^{\rm{I}}, \eta_{m,k}^{\rm{I}}\geq 0,\ \forall m,k$. 
	We define $\mv{A} =\sum_{k=1}^K \! p_k\!\sum_{m=1}^M\!\!\bigg(\!\frac{w\mv{a}_{m,k}\mv{a}_{m,k}^T}{s_{m,k}}\!+\!\frac{v_m\mv{a}_{m,k}\mv{a}_{m,k}^T}{b_{m,k}}\bigg)\!+\!\mv{F}_{\mathrm{P}}^{\mv{u}}\!\!$.
	Then, the Lagrangian of (P1-eqv-II) can be expressed as $
	\mathcal{L}_2(\{b_{m,k}\}, \{\mv{a}_{m,k}\}, \{\lambda_{m,k}^{\rm{I}}\}, \{\eta_{m,k}^{\rm{I}}\}) = {\rm{tr}}(
	\mv{A}^{-1})+\sum_{m=1}^{M}\sum_{k=1}^K\lambda_{m,k}^{\rm{I}}(\mv{a}_{m,k}^{T}\mv{a}_{m,k}-b_{m,k}^{\frac{1}{3}})+\sum_{m=1}^{M}\sum_{k=1}^K\eta_{m,k}^{\rm{I}}(\mv{a}_{m,k}^{T}\mv{a}_{m,k}-s_{m,k}^{\frac{1}{4}})$. 
	The partial derivative of the Lagrangian of (P1-eqv-II) with respect to $\{b_{m,k}\}$,  $\{s_{m,k}\}$, and $\{\mv{a}_{m,k}\}$, are given respectively by: 
	\begin{align}\label{L2_1}
		&\!\!\!\!\!\!\nabla_{{b}_{m,k}}{\mathcal{L}_2}(\{b_{m,k}\}, \{\mv{a}_{m,k}\}, \{{\lambda_{m,k}^{\rm{I}}}\}, \{{\eta_{m,k}^{\rm{I}}}\})\! \nonumber\\
		&\!\!\!\!=\!\! \frac{v_mp_k}{b_{m,k}}{\mv{a}_{m,k}}^{T}({\mv{A}}^{-1})^{2}\mv{a}_{m,k}\!\!-\!\!\frac{1}{3}{\lambda_{m,k}^{\rm{I}}}b_{m,k}^{-\frac{2}{3}}, \ \forall m\!\in\!\mathcal{M}, \forall k\!\in\!\mathcal{K}.
	\end{align}
	\begin{align}\label{L2_12}
			&\!\!\!\!\!\!\nabla_{{s}_{m,k}}{\mathcal{L}_2}(\{b_{m,k}\}, \{\mv{a}_{m,k}\}, \{{\lambda_{m,k}^{\rm{I}}}\}, \{{\eta_{m,k}^{\rm{I}}}\})\! \nonumber\\
		&\!\!\!\!=\!\! \frac{wp_k}{s_{m,k}}{\mv{a}_{m,k}}^{T}({\mv{A}}^{-1})^{2}\mv{a}_{m,k}\!\!-\!\!\frac{1}{4}{\lambda_{m,k}^{\rm{I}}}s_{m,k}^{-\frac{3}{4}}, \ \forall m\!\in\!\mathcal{M}, \forall k\!\in\!\mathcal{K}.
	\end{align}
	\begin{align}\label{L2_2}
		&\nabla_{{\mv{a}}_{m,k}}{\mathcal{L}_2}(\{b_{m,k}\}, \{\mv{a}_{m,k}\}, \{{\lambda_{m,k}^{\rm{I}}}\}, \{{\eta_{m,k}^{\rm{I}}}\})\! \nonumber\\
		&=-2p_{k}\left(\frac{w}{s_{m,k}}+\frac{v_m}{b_{m,k}}\right)(\mv{A}^{-1})^2\mv{a}_{m,k}+2(\lambda_{m,k}^{\rm{I}}+\nonumber\\
		&\eta_{m,k}^{\rm{I}})\mv{a}_{m,k},\ \forall m\!\in\!\mathcal{M}, \forall k\!\in\!\mathcal{K}\vspace{-2mm}
	\end{align}
	
	Let $ \!\mv{\Lambda} \! \!= \! \! \begin{bmatrix}
		\mv{\Lambda}_{1}  \!& \!\mv{\Lambda}_2  \\
		\mv{\Lambda}_{2}^H \! & \! \mv{\Lambda}_3 \\
	\end{bmatrix}  \!\!\succeq \! \! \mv{0}$, $\{\lambda_{m,k}^{\rm{II}}\}$, $\{\eta_{m,k}^{\rm{II}}\}$ denote the dual variables associated with the constraints in (\ref{P3_1_constraint1}), (\ref{P3_constraint1}), and (\ref{P3_constraint2}) in (P1-eqv-III), respectively, where $\lambda_{m,k}^{\rm{II}}, \eta_{m,k}^{\rm{II}}\geq 0,\ \forall m,k$.  Thus, the Lagrangian of Problem (P1-eqv-III) can be expressed as $\mathcal{L}_3(\mv{T}, \{b_{m,k}\}, \{\mv{a}_{m,k}\}, \mv{\Lambda},  \{\lambda_{m,k}^{\rm{II}}\},  \{\eta_{m,k}^{\rm{II}}\}) = {\rm{tr}}(\mv{T}) - {\rm{tr}}\bigg(\mv{\Lambda}\begin{bmatrix}
		\mv{B} &\mv{I}_3  \\
		\mv{I}_3 & \mv{T} \\
	\end{bmatrix}\bigg)+\sum_{m=1}^{M}\sum_{k=1}^K\lambda_{m,k}^{\rm{II}}(\mv{a}_{m,k}^{T}\mv{a}_{m,k}-b_{m,k}^{\frac{1}{3}})+\sum_{m=1}^{M}\sum_{k=1}^K\eta_{m,k}^{\rm{II}}(\mv{a}_{m,k}^{T}\mv{a}_{m,k}-s_{m,k}^{\frac{1}{4}})$.
	Let $\mv{\Lambda}^{\star}$ denote the optimal dual variables to (P1-eqv-III). We define $ \mv{B}=\!\sum_{k=1}^K \! p_k\!\sum_{m=1}^M\!\!\bigg(\!\frac{w\mv{a}_{m,k}\mv{a}_{m,k}^T}{s_{m,k}}\!+\!\frac{v_m\mv{a}_{m,k}\mv{a}_{m,k}^T}{b_{m,k}}\bigg)\!+\!\mv{F}_{\mathrm{P}}^{\mv{u}}\!\!$. Under the KKT conditions, we have
\begin{align}\label{L3_KKT_2}
		{\rm{tr}}\bigg(\mv{\Lambda}^{\star}\begin{bmatrix}
			\mv{B}^{\star} &\mv{I}_3  \\
			\mv{I}_3 & \mv{T}^{\star} \\
		\end{bmatrix}\bigg) = 0
	\end{align}
	\begin{align}\label{L3_KKT_1}
		&\nabla_{\mv{T}}\mathcal{L}_3(\mv{T}^{\star}, \{b_{m,k}^{\star}\}, \{\mv{a}_{m,k}^{\star}\}, \mv{\Lambda}^{\star},  \{{\lambda_{m,k}^{\rm{II}}}^{\star}\},  \{{\eta_{m,k}^{\rm{II}}}^{\star}\}) \nonumber\\
		&= \mv{I}_3-\mv{\Lambda}_3^{\star} = \mv{0}.
	\end{align}
	\begin{align}\label{L3_KKT_3}
		\!\!\!\!\!\!\!&\nabla_{\mv{a}_{m,k}}\mathcal{L}_3(\mv{T}^{\star}, \{b_{m,k}^{\star}\}, \{\mv{a}_{m,k}^{\star}\}, \mv{\Lambda}^{\star},  \{{\lambda_{m,k}^{\rm{II}}}^{\star}\},  \{{\eta_{m,k}^{\rm{II}}}^{\star}\})  \!\! \nonumber\\
		&=-2p_{k}(\frac{w}{s_{m,k}^{\star}}+\frac{v_m}{b_{m,k}^{\star}})\mv{\Lambda}_1^{\star}\mv{a}_{m,k}^{\star}+2({\lambda_{m,k}^{\rm{II}}}^{\star}+{\eta_{m,k}^{\rm{II}}}^{\star})\mv{a}_{m,k}^{\star}\nonumber\\&=\! \mv{0}, \  \forall m\!\in\!\mathcal{M}, \forall k\!\in\!\mathcal{K}
	\end{align}
	\begin{align}\label{L3_KKT_4}
		\!\!\!\!\!\!\!&\nabla_{{b}_{m,k}}{\mathcal{L}_3}(\mv{T}^{\star}, \{b_{m,k}^{\star}\}, \{\mv{a}_{m,k}^{\star}\}, \mv{\Lambda}^{\star}, \{\lambda_{3,m,k}^{\star}\})\!\nonumber\\
		&=\!\! \frac{v_mp_k}{{b_{m,k}^{\star}}^2}{\mv{a}_{m,k}^{\star}}^{T}{\mv{\Lambda}_1^{\star}}\mv{a}_{m,k}^{\star}\!\!-\!\!\frac{1}{3}{\lambda_{m,k}^{\rm{II}}}^{\star}{b_{m,k}^{\star}}^{-\frac{2}{3}}
		=\! {0}, \  \forall m\!\in\!\mathcal{M}, \forall k\!\in\!\mathcal{K}.
	\end{align}
		\begin{align}\label{L3_KKT_5}
		\!\!\!\!\!\!\!&\nabla_{{s}_{m,k}}{\mathcal{L}_3}(\mv{T}^{\star}, \{b_{m,k}^{\star}\}, \{\mv{a}_{m,k}^{\star}\}, \mv{\Lambda}^{\star}, \{\lambda_{3,m,k}^{\star}\})\!\nonumber\\
		&=\!\! \frac{wp_k}{{s_{m,k}^{\star}}^2}{\mv{a}_{m,k}^{\star}}^{T}{\mv{\Lambda}_1^{\star}}\mv{a}_{m,k}^{\star}\!\!-\!\!\frac{1}{4}{\lambda_{m,k}^{\rm{II}}}^{\star}{s_{m,k}^{\star}}^{-\frac{3}{4}}
		=\! {0}, \  \forall m\!\in\!\mathcal{M}, \forall k\!\in\!\mathcal{K}.
	\end{align}
	Note that the variables $\mv{T}$ is constrained to satisfy $\mv{T}\succeq \mv{B}^{-1}$, where $ \mv{B}=\!\sum_{k=1}^K \! p_k\!\sum_{m=1}^M\!\!\bigg(\!\frac{w\mv{a}_{m,k}\mv{a}_{m,k}^T}{s_{m,k}}\!+\!\frac{v_m\mv{a}_{m,k}\mv{a}_{m,k}^T}{b_{m,k}}\bigg)\!+\!\mv{F}_{\mathrm{P}}^{\mv{u}}\!\!$. To achieve the minimum objective value, the optimal solution must satisfy $\mv{T}^{\star} = {\mv{B}^{\star}}^{-1}$. 
	According to KKT condition in (\ref{L3_KKT_1}), we have $\mv{\Lambda}_3^{\star} = \mv{I}_3$. Since $\mv{\Lambda}\succeq\mv{0}$ and $\begin{bmatrix}
		\mv{B}^{\star} &\mv{I}_3  \\
		\mv{I}_3 & \mv{T}^{\star} \\
	\end{bmatrix}\succeq\mv{0}$, according to (\ref{L3_KKT_2}), the complementary slackness condition in (\ref{L3_KKT_2}) implies $\mv{\Lambda}^{\star}\begin{bmatrix}
		\mv{B}^{\star} &\mv{I}_3  \\
		\mv{I}_3 & \mv{T}^{\star} \\
	\end{bmatrix} =\mv{0}$. Expanding the block matrix multiplication leads to the following equalities: $\mv{\Lambda}_{1}^{\star}\mv{B}^{\star}+\mv{\Lambda}_{2}^{\star}= \mv{0}$, $\mv{\Lambda}_{1}^{\star}+\mv{\Lambda}_{2}^{\star}{\mv{B}^{\star}}^{-1} = \mv{0}$, ${\mv{\Lambda}_{2}^{\star}}^H\mv{B}^{\star}+\mv{I}_3 = \mv{0}$, and ${\mv{\Lambda}_{2}^{\star}}^H+{\mv{A}^{\star}}^{-1} = \mv{0}$. We can obtain the optimal dual variable as $\mv{\Lambda}_1^{\star} = ({\mv{B}^{\star}}^{-1})^{2}$. Moreover, we obtain the optimal dual variables: ${\lambda_{m,k}^{\rm{II}}}^{\star} = \frac{3v_mp_k}{{b_{m,k}^{\star}}^{-\frac{4}{3}}}{\mv{a}_{m,k}^{\star}}^T(\mv{B}^{-1})^2\mv{a}_{m,k}^{\star}$, and ${\eta_{m,k}^{\rm{II}}}^{\star} = \frac{4wp_k}{{s_{m,k}^{\star}}^{-\frac{5}{4}}}{\mv{a}_{m,k}^{\star}}^T(\mv{B}^{-1})^2\mv{a}_{m,k}^{\star}$. To verify the consistency of a KKT point between (P1-eqv-III) and (P1-eqv-II), we substitute the optimal variables $\mv{a}_{m,k}^{\star}$ ${s}_{m,k}^{\star}$ ,and ${b}_{m,k}^{\star}$ for $m\in\mathcal{M}, k\in\mathcal{K}$ into (\ref{L2_1}), (\ref{L2_12}), and (\ref{L2_2}) of (P1-eqv-II). It follows that $\nabla_{{\mv{a}}_{m,k}}\mathcal{L}_2(\{b_{m,k}^{\star}\}, \{\mv{a}_{m,k}^{\star}\}, \{{\lambda_{m,k}^{\rm{I}}}^{\star}\}, \{{\eta_{m,k}^{\rm{I}}}^{\star}\})= \mv{0}$, $\nabla_{{b}_{m,k}}\mathcal{L}_2(\{b_{m,k}^{\star}\}, \{\mv{a}_{m,k}^{\star}\}, \{{\lambda_{m,k}^{\rm{I}}}^{\star}\}, \{{\eta_{m,k}^{\rm{I}}}^{\star}\}) =0$, and $\nabla_{{s}_{m,k}}\mathcal{L}_2(\{b_{m,k}^{\star}\}, \{\mv{a}_{m,k}^{\star}\}, \{{\lambda_{m,k}^{\rm{I}}}^{\star}\}, \{{\eta_{m,k}^{\rm{I}}}^{\star}\})=0$. Hence, we conclude that any KKT point of (P1-eqv-III) also satisfies the KKT conditions of (P1-eqv-II), and thus the two problems share the same set of KKT points.
	
	In the following, we investigate the correspondence between the KKT points of problems (P1-eqv-II) and (P1-eqv-I).
The Lagrangian function of (P1-eqv-I) is given by
$\mathcal{L}_1(\{\mv{a}_{m,k}\})= {\rm{tr}}\bigg(\!\bigg(\!\sum_{k=1}^Kp_k\!\sum_{m=1}^M\!\bigg(\!\frac{w\mv{a}_{m,k}\mv{a}_{m,k}^T}{(\mv{a}_{m,k}^T\mv{a}_{m,k})^4}\!\!+\!\!\frac{v_m\mv{a}_{m,k}\mv{a}_{m,k}^T}{(\mv{a}_{m,k}^T\mv{a}_{m,k})^3}\!\bigg)\!\!+\!\!\mv{F}_{\mathrm{P}}^{\mv{u}}\!\bigg)^{-1}\bigg)
$. The corresponding gradient with respect to $\{\mv{a}_{m,k}\}$ is given by:
\begin{align}
&\!\nabla_{\mv{a}_{m,k}}\mathcal{L}_1(\{\mv{a}_{m,k}\}) = -p_k\left(\frac{6w}{
(\mv{a}_{m,k}^T\mv{a}_{m,k})^4}+\frac{4v_m}{
(\mv{a}_{m,k}^T\mv{a}_{m,k})^3}\right)\nonumber\\
&\!\bigg(\!\sum_{k=1}^Kp_k\!\sum_{m=1}^M\!\bigg(\!\frac{w\mv{a}_{m,k}\mv{a}_{m,k}^T}{(\mv{a}_{m,k}^T\mv{a}_{m,k})^4}\!\!+\!\!\frac{v_m\mv{a}_{m,k}\mv{a}_{m,k}^T}{(\mv{a}_{m,k}^T\mv{a}_{m,k})^3}\!\bigg)\!\!+\!\!\mv{F}_{\mathrm{P}}^{\mv{u}}\!\bigg)^{-1}\mv{a}_{m,k}\mv{a}_{m,k}^T\nonumber\\
&\!\bigg(\!\sum_{k=1}^Kp_k\!\sum_{m=1}^M\!\bigg(\!\frac{w\mv{a}_{m,k}\mv{a}_{m,k}^T}{(\mv{a}_{m,k}^T\mv{a}_{m,k})^4}\!\!+\!\!\frac{v_m\mv{a}_{m,k}\mv{a}_{m,k}^T}{(\mv{a}_{m,k}^T\mv{a}_{m,k})^3}\!\bigg)\!\!+\!\!\mv{F}_{\mathrm{P}}^{\mv{u}}\!\bigg)^{-1}\mv{a}_{m,k}.
\end{align}
	Since the constraints in (\ref{P3_constraint1}) and (\ref{P3_constraint2}) are tight at the optimum, it follows that $\mv{b}_{m,k}^{\star} = (\mv{a}_{m,k}^{\star}{\mv{a}_{m,k}^{\star}}^T)^3$, $\mv{s}_{m,k}^{\star} = (\mv{a}_{m,k}^{\star}{\mv{a}_{m,k}^{\star}}^T)^4,\forall m\in\mathcal{M}, \forall k\in\mathcal{K}$. By substituting the optimal solution $\mv{a}_{m,k}^{\star}$ obtained from (P1-eqv-II) into the Lagrangian function of (P1-eqv-I),  it can be verified that all KKT conditions of (P1-eqv-I) are satisfied, i.e., $\nabla_{\mv{a}_{m,k}}\mathcal{L}_1(\{\mv{a}_{m,k}^{\star}\}) = \mv{0}, \forall m,k$.
	Hence, the KKT point $\{\mv{a}_{m,k}^{\star}\}$ obtained from (P1-eqv-III) is also a KKT point for both (P1-eqv-II) and (P1-eqv-I), which completes the proof of Proposition~\ref{KKT_point}.

	\bibliographystyle{IEEEtran}
	
	\bibliography{IEEE-conference-template} 

\begin{thebibliography}{10}
\providecommand{\url}[1]{#1}
\csname url@samestyle\endcsname
\providecommand{\newblock}{\relax}
\providecommand{\bibinfo}[2]{#2}
\providecommand{\BIBentrySTDinterwordspacing}{\spaceskip=0pt\relax}
\providecommand{\BIBentryALTinterwordstretchfactor}{4}
\providecommand{\BIBentryALTinterwordspacing}{\spaceskip=\fontdimen2\font plus
\BIBentryALTinterwordstretchfactor\fontdimen3\font minus
  \fontdimen4\font\relax}
\providecommand{\BIBforeignlanguage}[2]{{%
\expandafter\ifx\csname l@#1\endcsname\relax
\typeout{** WARNING: IEEEtran.bst: No hyphenation pattern has been}%
\typeout{** loaded for the language `#1'. Using the pattern for}%
\typeout{** the default language instead.}%
\else
\language=\csname l@#1\endcsname
\fi
#2}}
\providecommand{\BIBdecl}{\relax}
\BIBdecl

\bibitem{liu2022integrated}
F.~Liu, Y.~Cui, C.~Masouros, J.~Xu, T.~X. Han, Y.~C. Eldar, and S.~Buzzi,
  ``Integrated sensing and communications: Toward dual-functional wireless
  networks for {6G} and beyond,'' \emph{IEEE J. Sel. Areas Commun.}, vol.~40,
  no.~6, pp. 1728--1767, Jun. 2022.

\bibitem{CM}
L.~Liu, S.~Zhang, , and S.~Cui, ``Leveraging a variety of anchors in cellular
  network for ubiquitous sensing,'' \emph{IEEE Commun. Mag.}, vol.~62, no.~9,
  pp. 98--104, Sep. 2024.

\bibitem{shi2022device}
Q.~Shi, L.~Liu, S.~Zhang, and S.~Cui, ``Device-free sensing in {OFDM} cellular
  network,'' \emph{IEEE J. Sel. Areas Commun.}, vol.~40, no.~6, pp. 1838--1853,
  Jun. 2022.

\bibitem{godrich2010target}
H.~Godrich, A.~M. Haimovich, and R.~S. Blum, ``Target localization accuracy
  gain in {MIMO} radar-based systems,'' \emph{IEEE Trans. Inf. Theory},
  vol.~56, no.~6, pp. 2783--2803, Jun. 2010.

\bibitem{nguyen2016optimal}
N.~H. Nguyen and K.~Do{\u{g}}an{\c{c}}ay, ``Optimal geometry analysis for
  multistatic {TOA} localization,'' \emph{IEEE Trans. Signal Process.},
  vol.~64, no.~16, pp. 4180--4193, Aug. 2016.

\bibitem{fatima2024optimal}
G.~Fatima, P.~Stoica, A.~Aubry, A.~De~Maio, and P.~Babu, ``Optimal placement of
  the receivers for multistatic target localization,'' \emph{IEEE Trans. Radar
  Syst.}, vol.~2, no.~6, pp. 391--403, Mar. 2024.

\bibitem{aubry2023robust}
A.~Aubry, P.~Babu, A.~De~Maio, G.~Fatima, and N.~Sahu, ``A robust framework to
  design optimal sensor locations for {TOA} or {RSS} source localization
  techniques,'' \emph{IEEE Trans. Signal Process.}, vol.~71, pp. 1293--1306,
  Mar. 2023.

\bibitem{xu2023mimo}
C.~Xu and S.~Zhang, ``{MIMO} radar transmit signal optimization for target
  localization exploiting prior information,'' in \emph{Proc. IEEE Int. Symp.
  Inf. Theory (ISIT)}, Jun. 2023, pp. 310--315.

\bibitem{xu2024mimo}
------, ``{MIMO} integrated sensing and communication exploiting prior
  information,'' \emph{IEEE J. Sel. Areas Commun.}, vol.~42, no.~9, pp.
  2306--2321, Sep. 2024.

\bibitem{bibTrees}
H.~L. Van~Trees, \emph{{Detection, Estimation, and Modulation Theory: Part I}},
  Wiley, New York, 1968.

\bibitem{Hou_JSAC}
K.~Hou and S.~Zhang, ``Optimal beamforming for secure integrated sensing and
  communication exploiting target location distribution,'' \emph{IEEE J. Sel.
  Areas Commun.}, vol.~42, no.~11, pp. 3125--3139, Nov. 2024.

\bibitem{skolnik1980introduction}
M.~I. Skolnik, \emph{Introduction to {Radar Systems}}.\hskip 1em plus 0.5em
  minus 0.4em\relax McGraw-Hill, New York, 1980.

\bibitem{marks1978general}
B.~R. Marks and G.~P. Wright, ``A general inner approximation algorithm for
  nonconvex mathematical programs,'' \emph{Operations research}, vol.~26,
  no.~4, pp. 681--683, 1978.

\bibitem{bibConvexOpt}
S.~Boyd and L.~Vandenberghe, \emph{{Convex Optimization}}.\hskip 1em plus 0.5em
  minus 0.4em\relax Cambridge, U.K.: Cambridge Univ. Press, 2004.

\bibitem{cvx}
M.~Grant and S.~Boyd, ``{CVX: Matlab software for disciplined convex
  programming},'' version 2.1. [Online]. Available: \url{http://cvxr.com/cvx/},
  Jun. 2015.

\bibitem{beck2015weiszfeld}
A.~Beck and S.~Sabach, ``Weiszfeld's method: Old and new results,'' \emph{J.
  Optim. Theory Appl.}, vol. 164, pp. 1--40, 2015.

\end{thebibliography}
\vspace{-3mm}
\end{document}